\documentclass[10pt,journal,twoside,web]{ieeecolor}

\usepackage{etoolbox}
\makeatletter
\@ifundefined{color@begingroup}%
  {\let\color@begingroup\relax
   \let\color@endgroup\relax}{}%
\def\fix@ieeecolor@hbox#1{%
  \hbox{\color@begingroup#1\color@endgroup}}
\patchcmd\@makecaption{\hbox}{\fix@ieeecolor@hbox}{}{\FAILED}
\patchcmd\@makecaption{\hbox}{\fix@ieeecolor@hbox}{}{\FAILED}

\usepackage{T_IEEEjournal}

\usepackage{amsmath,amssymb}
\usepackage{bm}
\usepackage{mathtools}

\newtheorem{theorem}{Theorem}
\newtheorem{assum}{Assumption}
\newtheorem{lemma}{Lemma}

\usepackage{graphicx}

\usepackage{tabularx}

\usepackage[hang,small,bf]{caption}
\usepackage[subrefformat=parens]{subcaption}
\usepackage[hyphens]{url}

\usepackage{comment}
\usepackage{cite}

\allowdisplaybreaks

\renewcommand{\QED}{\QEDopen}

\usepackage[dvipsnames]{xcolor}
\usepackage{booktabs,multirow}

\begin{document}
\title{Resilient Distribution Network Planning against Dynamic Malicious Power Injection Attacks}

\author{Hampei~Sasahara,~\IEEEmembership{Member},
        Tatsuya~Yamada,
        Jun-ichi~Imura,~\IEEEmembership{Senior Member},\\
        and Henrik Sandberg,~\IEEEmembership{Fellow}
\thanks{H. Sasahara is with the Department of Information Physics and Computing, Graduate School of Information Science and Technology, the University of Tokyo, Tokyo, 113-8656 Japan; e-mail: hsasahara@g.ecc.u-tokyo.ac.jp.} 
\thanks{T. Yamada, and J. Imura are with the Department of Systems and Control Engineering, School of Engineering, Institute of Science Tokyo, Tokyo, 152-8552 Japan;
e-mail: yamada@cyb.sc.e.titech.ac.jp, imura@sc.eng.isct.ac.jp.}
\thanks{H. Sandberg is with Division of Decision and Control Systems, KTH Royal Institute of Technology, Stockholm, SE-100 44 Sweden; e-mail: hsan@kth.se.}
\thanks{This work was supported in part by JSPS KAKENHI Grant Number JP24K17296 and the Swedish Research Council grant 2023-04770.}
\thanks{Manuscript received Xxx xx, 20xx; revised Xxx xx, 20xx.}
\thanks{This is the authors' version of an accepted paper. The final published version is available at IEEE Xplore.}
}

\markboth{IEEE Transactions on Control of Network Systems,~Vol.~XX, No.~XX, November~2025}%
{Sasahara \MakeLowercase{\textit{et al.}}: Resilient Distribution Network Planning against Dynamic Malicious Power Injection Attacks}

\maketitle

\begin{abstract}
Active distribution networks facilitating bidirectional power exchange with renewable energy resources are susceptible to cyberattacks due to integration of a diverse array of cyber components.
This study introduces a grid-level defense strategy aimed at enhancing attack resiliency based on distribution network planning.
Our proposed framework imposes a security requirement into existing planning methodologies, ensuring that voltage deviation from its rated value remains within a tolerable range against dynamically and maliciously injected power at end-user nodes.
Unfortunately, the formulated problem in its original form is intractable because it is an infinite-dimensional bi-level optimization problem over a function space.
To address this complexity, we develop an equivalent transformation into a tractable form as mixed-integer linear program leveraging linear dynamical system theory and graph theory.
Notably, our investigation reveals that the severity of potential attacks hinges solely on the cumulative reactances over the path from the substation to the targeted node, thereby reducing the problem to a finite-dimensional problem.
Further, the bi-level optimization problem is reduced to a single-level optimization problem by using a technique utilized in solving the shortest path problem.
Through extensive numerical simulations conducted on a 54-node distribution network benchmark, our proposed methodology exhibits a noteworthy 29.3\% enhancement in the resiliency, with a mere 2.1\% uptick in the economic cost.
\end{abstract}

\begin{IEEEkeywords}
Distribution network planning, power injection attack, resilient distribution systems.
\end{IEEEkeywords}

%
\IEEEpeerreviewmaketitle

\section{Introduction}
%
%
%
%
\IEEEPARstart{M}{assive} deployment of renewable energy resources, such as solar and wind power, has triggered a significant change in the traditional manual approach to planning and operating passive distribution networks, resulting in the emergence of active distribution networks (ADNs) that facilitate bidirectional power exchange~\cite{Ehsan2019State}.
Contemporary advancements such as the Internet of Things (IoT) and cloud computing serve as the building blocks of ADNs, playing a role of timely suppression of voltage fluctuations, efficient control of power flows, and strong enhancement of reliability and stability of the grid under the bidirectional flows.
On the other hand, the rush to integration of a wide variety of cyber components with ADNs aggravates the risk of cyberattacks~\cite{Khalaf2024Survey}.
In particular, the dynamic nature of ADNs, characterized by frequent reconfigurations and the need for decentralized decision-making, necessitates advanced cybersecurity frameworks that can adapt to evolving threats~\cite{Tesfay2014Cyber}.
For instance, the adoption of software-defined networking has been proposed to enhance communication security and ensure the integrity of decentralized energy management processes~\cite{Li2018Cyber}.
In the real world, in 2015 cyber attack on the Ukrainian power system started at the distribution level and results in a prolonged power outage affecting over 225,000 customers~\cite{Liang2017Ukraine}.
The threat to power systems continued and two severe attacks have been reported in 2016 and 2022~\cite{Salazar2024Tale}.
As demonstrated by those attacks, designing resilient distribution grids against cyberattacks is now a crucial issue of ADNs.

In particular, edge devices at the end-user level are liable to cyberattacks because of the absence of standardization~\cite{Husnoo2023False}.
A typical example of the newly emerging attack surfaces at the end-user level is modern grid-tied smart inverters~\cite{Li2022Cybersecurity}, whose components are electrical devices such as DC/DC converter, DC/AC inverter, and low frequency (LF) transformer, and control devices such as micro control unit (MCU), digital signal processor (DSP), and communication module.
Software architects prefer to base their designs in existing solutions for handling the complexity and diversity, and the off-the-shelf integration tends to leave the system vulnerable.
Indeed, it has been reported that a few dozen of vulnerabilities of smart inverter software for photovoltaic (PV) systems can be found in public databases~\cite{Dubasi2021Security}.
While component-level cybersecurity measures~\cite{Qi2016Cybersecurity,Xu2023Resilience}, play a critical role in protecting individual components of power distribution networks, they are insufficient due to the high degree of interdependency among devices.
A vulnerability in one device can impact the performance or security of others, meaning that even a well-protected component may be affected by the compromise of another.
Component-level defenses are unable to account for these complex interactions that emerge from the behavior of multiple devices~\cite{Krause2021Cybersecurity}.
Additionally, maintaining consistent security policies across a large and diverse set of devices is operationally difficult~\cite{Sou2024Resilient}.
Indeed, recent real-world findings, such as the discovery of unauthorized communication modules in smart inverters reported in 2025~\cite{Mcfarlane2025Rogue}, demonstrate that adversaries can potentially exploit such embedded components to compromise system-wide operations, underscoring the urgency of proactive, grid-level defense mechanisms.
Grid-level cybersecurity is therefore essential, as it provides system-wide visibility and centralized policy enforcement.
By addressing the risks that arise from device interdependence, grid-level defenses help ensure the resilience, integrity, and reliable operation of the entire distribution network.

This study introduces a novel approach to distribution network planning aimed at fortifying grid-level resiliency against cyberattacks.
The distribution network planning problem, encompassing both distribution system expansion and reconfiguration, represents a longstanding focal point within this domain.
In essence, planning entails strategic installation or reconfiguration of electric equipment to optimally meet technical specifications in a cost-efficient manner~\cite{Lavorato2012Imposing}.
Traditionally, this problem is formulated as an optimization problem, typically in the form of the mixed-integer linear program (MILP)~\cite{Rastgou2024Distribution}.
Based on existing planning frameworks, we impose an attack-resiliency criterion necessitated by the disruptive attack scenario that maximizes potential impact.
In particular, we treat \emph{dynamic malicious power injection} as a unified model of the grid-level effect of potential cyberattacks.
This abstraction proves particularly apt in scenarios where cyberattacks manipulate the behavior of smart inverters, exerting full control over their dynamics.
Further, we adopt \emph{voltage volatility}~\cite{Song2019New} as the key metric for assessing resiliency.
This metric serves a dual purpose, not only facilitating the prevention of protection relay tripping but also furnishing a mathematical framework for articulating the resiliency requirement.
Specifically, we constrain voltage fluctuations from their rated value, ensuring that they remain within predefined tolerances across the entire time horizon.

Imposing the security constraint into the distribution network planning problem poses two significant technical challenges.
Firstly, it becomes an infinite-dimensional optimization problem over a function space, since the attack signal dynamically influences the grid's behavior over time.
Secondly, it becomes a bi-level optimization problem, wherein the maximally disruptive power injection signal determined by the attacker depends on the network configuration determined by the grid planner.
To surmount the first challenge, we characterize the maximally disruptive power injection signal leveraging insights from linear dynamical system theory.
Our analysis reveals a notable monotonicity in radial grid's dynamics, affording us an analytical representation of the maximally disruptive power injection signal.
This characterization underscores that the severity of the attack impact escalates with the cumulative reactances of the lines over the path from the substation to the targeted node.
This pivotal discovery allows us to compress the original infinite-dimensional optimization problem, wherein the attacker decides both of the attack function and the targeted node, into a finite-dimensional counterpart, where the attacker solely determines the targeted node.
The key observation to overcome the second challenge is that the reduced optimization problem aligns with a variant of the bottleneck shortest path problem~\cite{Turner2011Variants}.
Then the bi-level optimization problem can further be reduced to a single-level optimization problem by the application of graph theory techniques~\cite{Rafael2016Formulations}.
This equivalent transformation renders the original intractable problem into an MILP, which is tractable using standard solvers.
We conduct numerical simulations using the widely utilized 54-node distribution network benchmark~\cite{Lavorato2012Imposing,Cruz2021Model,Miranda1994GA} with PV systems.
Our simulations reveal a remarkable 29.3\% enhancement in resiliency, achieved with a mere 2.1\% increase in economic cost.

Our contributions can be summarized as follows:
We introduce a novel distribution planning methodology aimed at enhancing grid-level resiliency against cyberattacks.
Furthermore, we derive an equivalent transformation of the proposed optimization problem into a tractable form by leveraging linear dynamical system theory and graph theory.

\subsection*{Related Work}

Resilience against extreme events, ranging from natural disasters to cyberattacks, stands as a cornerstone in the design of robust distribution grids~\cite{Gholami2018Towards,Bhusal2020Power}.
Broadly resilience enhancement techniques in distribution grids can be classified into two categories: proactive planning approaches and reactive operational strategies~\cite{Dwivedi2024Technological}.
Although integrating both approaches synergistically fortifies distribution grids~\cite{Yuan2016Robust,Huang2017Integration}, our proposed method aligns with the planning paradigm.

Numerous studies have explored resilience-enhanced distribution network planning~\cite{Song2019New,Saberi2023Power,Zu2018Distribution,Byeon2020Communication,Zou2020Resilient,Saravi2022Resilience,Bayani2023Resilient}.
For example, the work~\cite{Saberi2023Power} proposes a distribution network expansion planning with PV and gas-fired sources to improve resilience against high-impact low-probability events,
and the authors of~\cite{Zu2018Distribution} present a network reconfiguration method ensuring adherence to the $N-1$ criterion, thereby maintaining power supplies under any single contingency.
However, the majority of these studies employ the amount of unmet demand in the grid as the resiliency metric, overlooking critical factors such as voltage volatility being essential for understanding protective relay behavior and pre-failure processes.
Notably, an exceptional work~\cite{Song2019New} considers voltage volatility as the resiliency metric.
Nonetheless, their model simplifies the contingency effect as a static fluctuation of distributed generation, failing to capture the dynamic effects of maliciously controlled power.
Similarly, although voltage stability based on loadability analysis has been extensively studied~\cite{Akbarzadeh2020Voltage}, the majority of existing studies adopt a static framework, in which the effects of steady-state or slowly varying load changes are analyzed.
In the security context, the presence of an intelligent adversary necessitates analysis of dynamic effects~\cite{Manandhar2014Detection,Liu2022Relentless}.
Indeed, our result reveals that time-varying dynamic characteristics can significantly influence the system’s response.

Our threat with malicious power injection attack is distinguished from faults by its dynamic nature and the active involvement of a hostile entity~\cite{Gao2015Survey}.
A few studies can be found on grid-level resiliency of distribution networks against cyberattacks~\cite{Shelar2017Security,Shelar2021Evaluating,Giglou2022igdt,Ghanbari2023Resilient}.
The study~\cite{Shelar2017Security} addresses security assessment under distributed energy resource node compromises.
The work~\cite{Shelar2021Evaluating} offers a resiliency evaluation method under reactive measures based on a bi-level optimization and proposes a computational framework using Benders decomposition.
While they provide insights into evaluation, they do not present any planning methodology for resiliency enhancement.
A planning approach using dynamic programming based on information gap decision theory is proposed in~\cite{Giglou2022igdt}, where they consider false data injection attack on price information through smart meters. 
The study~\cite{Ghanbari2023Resilient} proposes an operation approach utilizing renewable energy and battery energy storage under an abstract model of cyberattacks.
In contrast, our work represents a pioneering effort in providing a planning approach for improving voltage volatility under realistic threat scenarios that impact dynamic behavior of distribution generation.
By addressing the dynamic effects of malicious power injection, our methodology fills a crucial gap in the existing literature on grid-level resilience enhancement against cyberattacks.

The optimization problem addressed in this study is fundamentally an infinite-dimensional bi-level program.
Traditional optimal control theory has provided powerful tools for analyzing infinite-dimensional problems, notably through frameworks such as Pontryagin’s Maximum Principle~\cite{Liberzon2011Calculus}.
However, these methods are predominantly applicable to unconstrained settings.
For constrained infinite-dimensional problems, standard numerical approaches include discretization-based methods~\cite{Hinze2011Discretization} and function approximation techniques~\cite{Devolder2010Solving}.
While these methods offer practical means of computation, they are often hampered by the curse of dimensionality and suffer from inherent approximation errors.
In contrast, our approach addresses the infinite-dimensional nature of the problem by exploiting the monotonicity properties in distribution networks.
This allows us to reformulate the problem in a more tractable manner without resorting to high-dimensional discretizations.
Bi-level optimization problems are generally known to be challenging, owing to their intrinsic nonconvexity and nonlinearity~\cite{Colson2007Overview}.
Nonetheless, several studies have proposed efficient algorithms by leveraging specific structural properties of the problem at hand~\cite{Chen2017Stackelberg}.
In our case, we capitalize on a key insight that the total reactance along a path from the substation has a direct and monotonic influence on the severity of an attack.
This insight enables us to reformulate the original problem as an MILP, significantly enhancing its tractability.

\subsection*{Organization}
The structure of this paper is outlined as follows:
In Sec.~\ref{sec:problem}, we formally provide the distribution network model, define the attack model, and formulate the resilient distribution network planning problem against dynamic malicious power injection attack.
Sec.~\ref{sec:assessment} treats the maximally disruptive power injection signal characterization problem as a subproblem and derives its analytic solution.
Based on the characterization we propose an equivalent transformation of the original problem into a tractable form in Sec.~\ref{sec:milp}.
Sec.~\ref{sec:simulation} verifies the effectiveness of the proposed method through numerical simulation using the 54-node distribution network and Sec.~\ref{sec:conclusion} draws the conclusion.

\section{Modeling and Problem Description}
\label{sec:problem}

\subsection{Distribution network components and structure}

\begin{figure}
    \centering
    \includegraphics[width=0.9\linewidth]{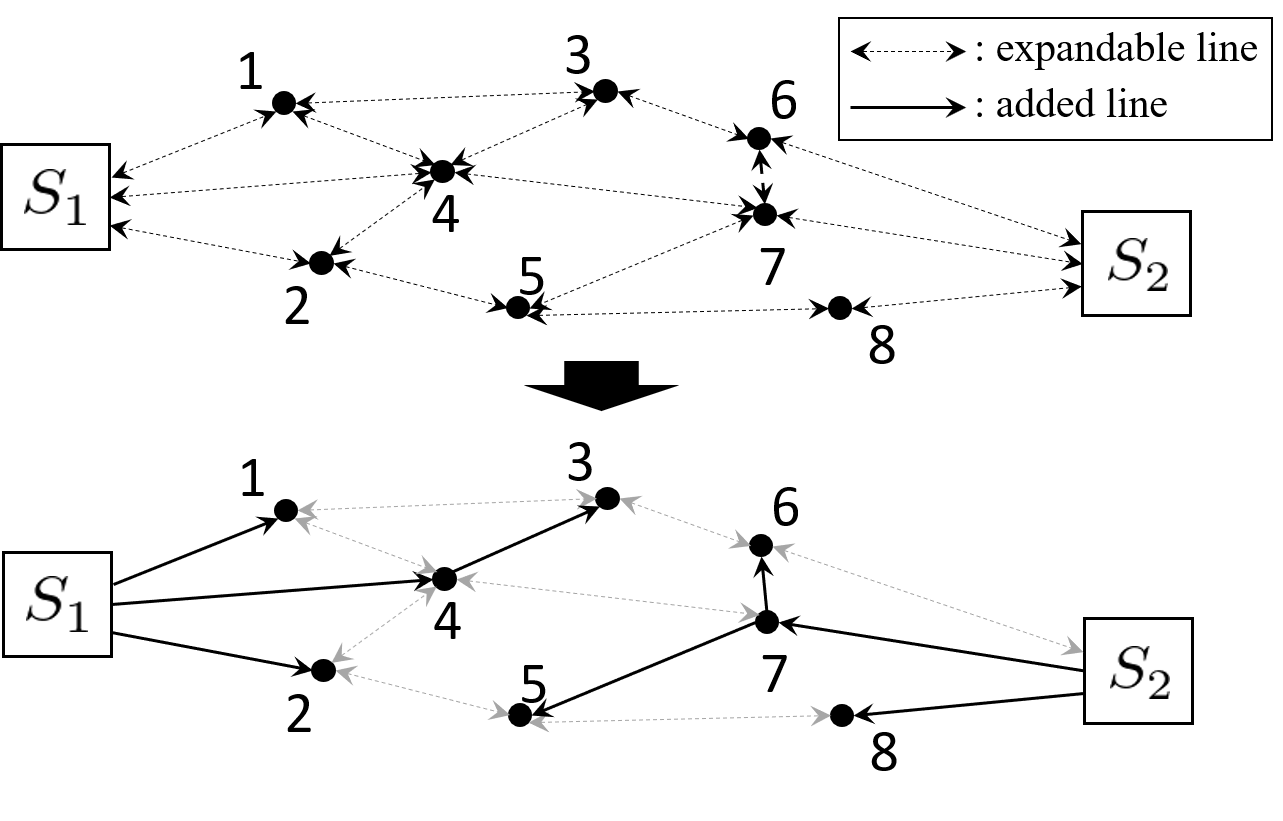}
    \caption{Illustration of distribution network planning.}
    \label{fig:ex-design}
\end{figure}

The power distribution network planning problem involves designing and optimizing the infrastructure needed to deliver electricity from the transmission system to end-users in the most efficient, reliable, and cost-effective manner.
Given an initial topology, the planner can install and reinforce electronic facilities, such as substations, conductors, circuits, and distributed energy resources, taking into account the load balance and radiality of the network.
Typical objective functions comprise annual investment and operation cost~\cite{Lavorato2012Imposing}, often incorporating reliability indices as well~\cite{Aschidamini2022ExpansionPlanning}.
In this paper, we consider only addition of lines as the decision variables, which is illustrated by Fig.~\ref{fig:ex-design}.
Further, we focus solely on the investment cost as the objective function for simplicity.
However, it should be noted that our formulation can readily be extended to accommodate more general scenarios by imposing our proposed attack-resiliency constraint.

\begin{figure}
    \centering
    \includegraphics[width=0.95\linewidth]{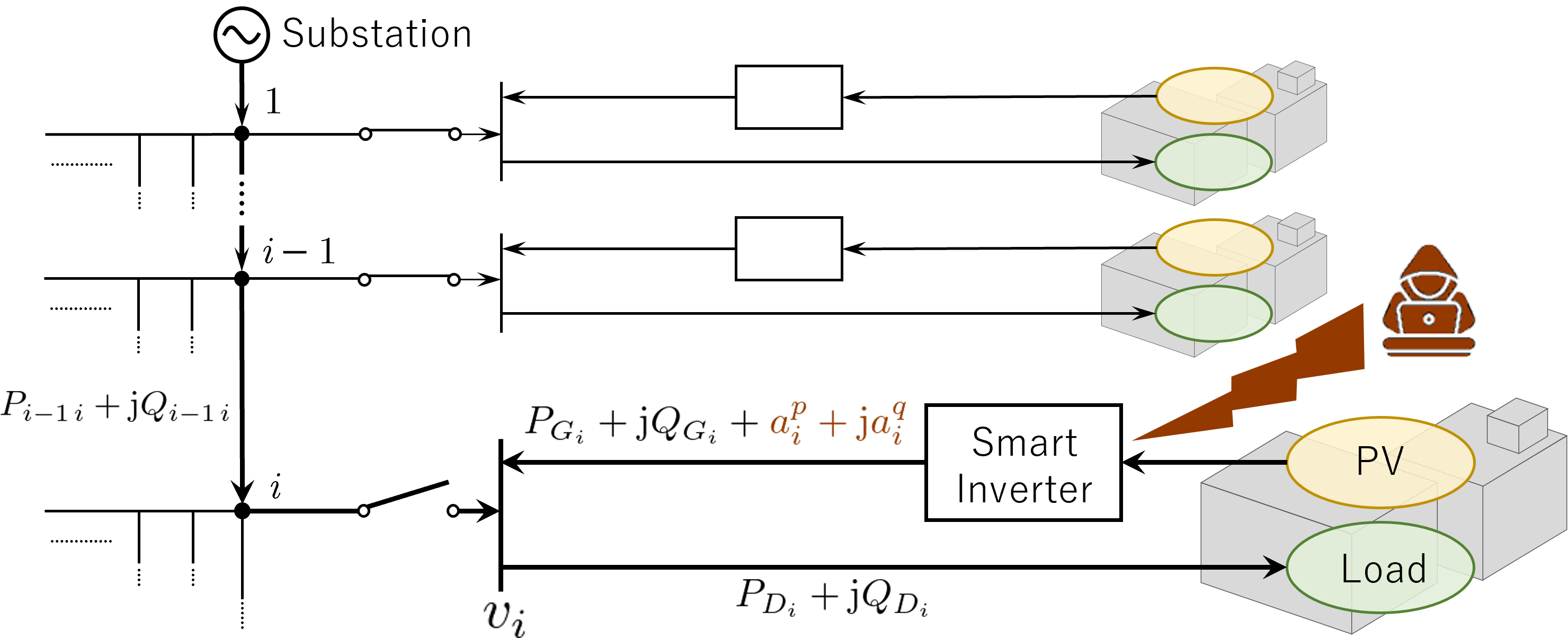}
    \caption{Consumer node structure and malicious power injection attack.}
    \label{fig:relay}
\end{figure}

We treat the underlying topology of the distribution network with PV systems as a directed graph.
Let $\Omega_{\rm N}$ and $\Omega_{\rm E}$ denote the sets of nodes and usable edges, respectively.
Each node is classified into a substation node or a consumer node, i.e.,
$\Omega_{\rm N}=\Omega_{\rm S} \cup \Omega_{\rm L}$, which denote the disjoint sets of substation nodes and consumer nodes, respectively.
While each substation node has a substation that supplies power to the grid,
each consumer node has an individual consumer with a load and a PV system as shown in Fig.~\ref{fig:relay}.
We disregard transmission nodes, in other words, each node is connected to either a substation or a consumer.
The direction at each edge describes its reference direction.

Topological requirements for our planning problem are as follows: (i) there exists at most a single directed edge per branch; (ii) the demand of every consumer node is met by the connected substations;
(iii) each subgraph has a radial structure.
The first requirement is described by
\begin{equation}\label{eq:oneway}
 n_{ij}+n_{ji}\leq 1\quad \forall (ij)\in \Omega_{\rm E}.
\end{equation}
Eq.~\eqref{eq:oneway} simply means that there is no bidirectional branch, where $n_{ij}$ is the binary decision variable which takes $1$ when edge $(ij)$ is used and $0$ otherwise.
We here introduce the linearized dist-flow model~\cite{Baran1989Optimal} where the power flow and voltage drop over each edge are expressed in a linear form.
Then the second requirement is described by
\begin{align}
    &P_{G_i}-P_{D_i}-\textstyle{\sum_{j\in\Omega_{\rm N}}P_{ij}=0} && \forall i\in\Omega_{\rm L}, \label{eq:Pflow} \\
    &Q_{G_i}-Q_{D_i}-\textstyle{\sum_{j\in\Omega_{\rm N}}Q_{ij}=0} && \forall i\in\Omega_{\rm L}, \label{eq:Qflow} \\
    &P_{S_i}-\textstyle{\sum_{j\in\Omega_{\rm N}}n_{ij}P_{ij}=0} && \forall i\in\Omega_{\rm S}, \label{eq:Psub} \\
    &Q_{S_i}-\textstyle{\sum_{j\in\Omega_{\rm N}}n_{ij}Q_{ij}=0} && \forall i\in\Omega_{\rm S}, \label{eq:Qsub} \\
    &0\leq P_{ij}\leq n_{ij}M && \forall(ij)\in\Omega_{\rm E}, \label{eq:Pconstraint} \\
    &0\leq Q_{ij}\leq n_{ij}M && \forall(ij)\in\Omega_{\rm E}, \label{eq:Qconstraint} \\
    &v^2_i-v^2_j=2(R_{ij}P_{ij}+X_{ij}Q_{ij}) && \forall (ij)\in\Omega_{\rm E} \label{eq:vol_drop}.
\end{align}
Eqs.~\eqref{eq:Pflow} and~\eqref{eq:Qflow} describe the power balance that must be met at each consumer node, where $P_{G_i}$ and $P_{D_i}$ denote given active powers generated and consumed at node $i$, respectively, and $P_{ij}$ denotes the active power flowing through edge $(ij)$.
The reactive powers $Q_{G_i}, Q_{D_i}$ and $Q_{ij}$ are defined in a similar manner.
Eqs.~\eqref{eq:Psub} and~\eqref{eq:Qsub} describe the power flow supplied by each substation, where $P_{S_i}$ and $Q_{S_i}$ denote the active and reactive powers supplied from substaition $i$, respectively.
Eqs.~\eqref{eq:Pconstraint} and \eqref{eq:Qconstraint} describe that power does not flow between two nodes when there is no edge between them, where $M$ is a large constant.
Eq.~\eqref{eq:vol_drop} describes the voltage drop over each branch, where $v_i$ denotes the voltage magnitude at node $i$ and $R_{ij}$ and $X_{ij}$ denote the given resistance and reactance of edge $(ij)$, respectively.
Under the power flow constraint, the third requirement is described by
\begin{equation}\label{eq:radiality}
\textstyle{
    \sum_{(ij)\in\Omega_E}n_{ij}=N_b-N_{b_S},
    }
\end{equation}
where $N_b$ and $N_{b_S}$ denote the numbers of all nodes and substation nodes, respectively.
Eq.~\eqref{eq:radiality} guarantees the radial structure of each subgrid~\cite{Lavorato2012Imposing}.
The reference directions are defined from the corresponding substation toward its downstream nodes, which determines the orientation of the associated binary decision variables $n_{ij}$ accordingly.
Fig.~\ref{fig:ex-design} exemplifies a planning of a distribution network, resulting in two subnetworks consisting of nodes $S_1,1,2,3,4$ and $S_2,5,6,7,8$, respectively, where $\Omega_{\rm L}=\{1,2,\ldots,8\}, \Omega_{\rm S} = \{S_1,S_2\},$ and $n_{S_1 1}=n_{S_1 2}=n_{S_1 4}=n_{43}=n_{75}=n_{76}=n_{S_2 7}=n_{S_2 8}=1$.

Subsequently, we provide a dynamical model of each PV system with a smart inverter that manipulates reactive power for voltage control.
We here consider a quadratic droop controller with an integrator~\cite{Chong2019Local,Simpson2016Voltage} as a model of the inverter, whose dynamics is given by
\begin{equation}\label{eq:inverter}
    \dot{Q}_{G_i}(t)=-K_i(v^2_i(t)-v^2_0(t)),
\end{equation}
which has a servomechanism that regulates the voltage to its rated value,
where $K_i$ denotes the given integral coefficient and $v_0$ denotes the given rated voltage magnitude.
Note that we here express the time argument $t$ to emphasize that this is a dynamical system described by a differential equation.
In conventional planning problems, distribution networks are treated as a static system because its dynamics is negligibly fast and all physical values converge to steady state quickly.
However, our interest is in analyzing effects of dynamic malicious power injection, introduced later, and hence we need to treat the distribution network as a dynamical system.
This feature makes our problem challenging as observed in Sec.~\ref{subsec:problem}.

\subsection{Attack model, resiliency metric, and security requirements}

\begin{figure}
    \centering
    \includegraphics[width=0.98\linewidth]{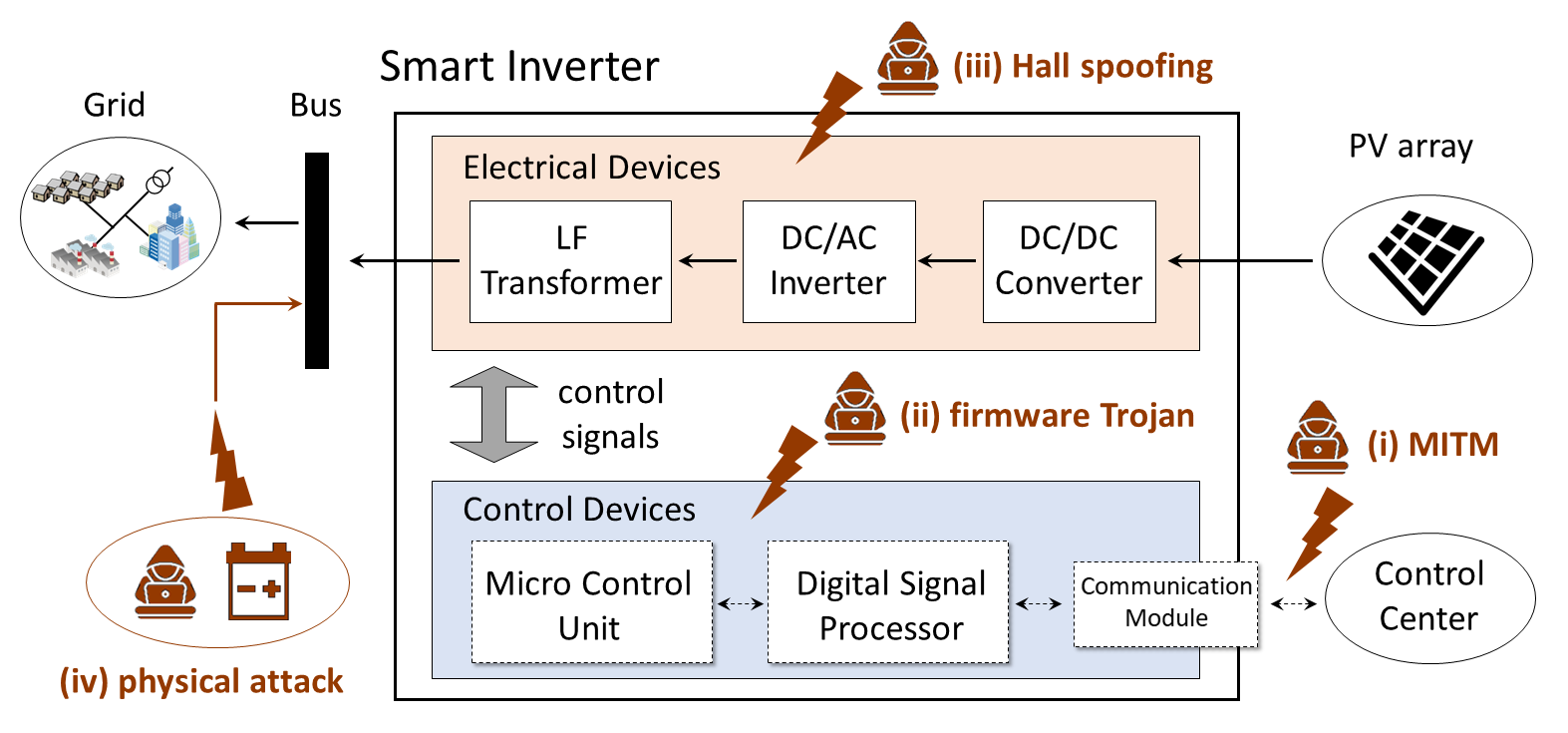}
    \caption{Architectural components of a smart inverter and possible threats.}
    \label{fig:smart}
\end{figure}

Grid-tied smart inverters play a pivotal role as the essential link between solar PV and the smart grid for boosting the overall performance by efficient control~\cite{Zeb2018Comprehensive}.
Fig.~\ref{fig:smart} illustrates the key architectural components of a smart inverter.
The power harvested from sunlight by the PV array is optimized using maximum power point tracking (MPPT) techniques within the DC/DC converter.
Subsequently, the DC/AC inverter and LF transformer convert the electricity to meet the grid's voltage, current, and frequency specifications.
The rectified power is then supplied to the connected facility and the grid via the connected bus.
Control devices including MCU and DSP are responsible for maintaining power quality by utilizing pulse width modulation (PWM) signals to achieve high efficiency, mitigate harmonic distortions, and regulate voltage levels.
Moreover, network communication between the client and the control center, facilitated by communication modules, enables seamless operation by supporting operator-dispatched settings that are heavily reliant on the grid status.
To realize these functionalities, the communication software commonly utilizes established components such as the Linux kernel, which encompasses networking stacks, file systems, software management applications, and remote access tools~\cite{Sebastian2019Tee}.

The proliferation of newly developed electrical and control devices with communication interfaces in smart inverters enlarges the attack surface in user-side PV systems~\cite{Li2022Cybersecurity}.
These vulnerabilities manifest in various forms and Fig.~\ref{fig:smart} describes four specific threats:
(i)~Man-in-the-Middle attack (MITM) intercepts and potentially alters communication between the client and the control center who believe that they are directly communicating with each other.
The attacker can eavesdrop on the communication, manipulate the data being transmitted, or even impersonate one or both parties involved~\cite{Carter2017Cyber}.
(ii)~Firmware Trojan is malicious software designed to infect and compromise the firmware of the control devices~\cite{Konstantinou2016Taxonomy}.
Firmware can be replaced by an attacker especially when manufacturers release their firmware files to the public~\cite{Sebastian2019Tee}.
(iii)~Hall spoofing targets the availability of inverters by spoofing Hall sensors in the electrical devices, manipulating the measurement signal sent to control devices by injecting false data~\cite{Barua2020Hall}.
Hall spoofing can be executed using distinct types of external magnetic
fields, even employing tools camouflaged within innocuous items such as flower vases.
(iv)~Physical attack directly intervenes in physical devices unlike intrusion into cyber components~\cite{Chung2018Local}.
These attacks, such as injecting extra power or consuming generated power by connecting generators or battery storage to the bus, can have severe consequences.
These vulnerabilities underscore the critical importance of robust security measures to safeguard smart inverters and mitigate potential threats in distribution networks.

Given the intricate and diverse nature of potential threats to PV systems, constructing a unified and detailed model of attacks poses considerable challenges.
In light of this complexity, we consider {\it additive dynamic malicious power injection} as a simplified mathematical model to offer a focused perspective on the grid-level impact of attacks~\cite{Adepu2020Attacks}.
Specifically, the net effect of the attacks is modeled as a perturbation to the output of the PV system.
The active power and reactive power flowing over the $i$th consumer node are perturbed by the attack as
\begin{equation}
\left\{
\begin{array}{l}
 P_i^{\rm a}(t)=P_{G_i}(t)-P_{D_i}(t)+a^{p}_i(t),\\
 Q_i^{\rm a}(t)=Q_{G_i}(t)-Q_{D_i}(t)+a^{q}_i(t),
\end{array}
\right.
\end{equation}
respectively, where $a^{p}_i(t)$ and $a^{q}_i(t)$ denote maliciously injected powers and $P_i^{\rm a}(t)$ and $Q_i^{\rm a}(t)$ denote the resulting power flows at the time instant $t$.
Again, we express the time argument to emphasize that this signal dynamically influences the grid's behavior.
The additive form of the injected active and reactive power significantly simplifies security analysis and makes the resilient distribution network planning problem tractable as demonstrated in Secs.~\ref{sec:assessment} and~\ref{sec:milp}.
The malicious power injection modifies~\eqref{eq:Pflow} and~\eqref{eq:Qflow} as
\begin{align}
    P_i^{\rm a} - \textstyle{\sum_{j\in\Omega_{\rm N}}P_{ij}=0} && \forall i\in\Omega_{\rm L}, \label{eq:Pflowr} \\
    Q_i^{\rm a}-\textstyle{\sum_{j\in\Omega_{\rm N}}Q_{ij}=0} && \forall i\in\Omega_{\rm L},\label{eq:Qflowr}
\end{align}
where~\eqref{eq:Pflowr} and~\eqref{eq:Qflowr} represent the modified power flow equations.
We describe the attacker's limited budget as a magnitude constraint on the apparent power such that $|a_i(t)|\leq C$ with some constant $C>0$ where $a_i(t):=a^p_i(t)+\sqrt{-1}a^q_i(t)$.
Although active and reactive power injections require distinctive control schemes, we impose this type of constraint because it significantly simplifies the analytical treatment as demonstrated in the subsequent section.
We describe the set of admissible attack signals by
$\mathcal{A}:=\{a_i \in \mathcal{L}^\infty: |a_i(t)|\leq C,\ \forall t\geq 0\}.$
Furthermore, we assume for simplicity that a single consumer node is targeted, i.e., $a_j(t)=0$ for any $t\geq 0$ for $j\neq i$ where $i$ is the index of the targeted node.
It should be emphasized that the dynamic nature of our threat model sets this work apart from existing studies.
For example, in~\cite{Shelar2021Evaluating} the attack consequence is abstractly modeled as a static disconnection of the target node.
In contrast, our attack model captures the dynamic consequences of maliciously controlled power designed to induce relay tripping.
This distinction highlights the detailed and temporally-evolving nature of the threat considered in our work, offering a more realistic and nuanced view of potential vulnerabilities.

As a resiliency metric against the dynamic malicious power injection, we employ \emph{voltage volatility}~\cite{Song2019New} motivated by the fact that large voltage fluctuation leads to activation of protective relays and disconnection of the customer node.
We pose the security requirement that the voltage at the targeted consumer node is kept within the tolerable range for any possible attacks.
The voltage volatility constraint is expressed by
$v_{\min}\leq v_i(t;a_i) \leq v_{\max}$
for any $t\in[0,+\infty)$ and $a_i\in\mathcal{A}$
with the given maximum and minimum tolerable voltage magnitudes $v_{\max}$ and $v_{\min}$, respectively,
where the argument $a_i$ is also expressed to emphasize that the voltage depends on the attack signal.
This inequality is equivalently rewritten as $v_{\min}^2 - v_0^2\leq v_i(t;a_i)^2-v_0^2 \leq v_{\max}^2-v_0^2.$
For simplicity, we here assume that $|v_{\max}^2-v_0^2|=|v_{\min}^2 - v_0^2|$ and then the constraint is described in the following form
\begin{equation}\label{eq:con_sec_trans}
\|y_i(\cdot\,;a_i)\|_{\infty} \leq \overline{y}\quad \forall a_i\in\mathcal{A}
\end{equation}
where $y_i(t;a_i):=v^2_i(t;a_i)-v^2_0$, $\overline{y}:=|v^2_{\max}-v^2_0|$, and $\|y_i(\cdot\,;a_i)\|_{\infty}:=\sup_{t\in[0,+\infty)}|y(t;a_i)|$ represent the squared voltage deviation, its tolerable range, and its $\mathcal{L}^{\infty}$ norm, respectively.
Since this form is compatible to the inverter dynamics~\eqref{eq:inverter} and easy to handle, we consider~\eqref{eq:con_sec_trans} hereinafter.

\subsection{Resilient distribution network planning problem}
\label{subsec:problem}

We formulate the resilient distribution network planning problem as follows:
\begin{align}
    \min \quad & J \notag \\
    {\rm s.t.}\quad &\eqref{eq:oneway}, \eqref{eq:Psub}\textrm{-}\eqref{eq:Qflowr}\\
    & \sup_{a_i\in\mathcal{A}}||y_i(\cdot\,;a_i)||_\infty \leq \overline{y}\quad \forall i\in\Omega_{\rm L}.\label{eq:sec_con}
\end{align}
with decision variables $n_{ij},v^2_i,P_{ij},Q_{ij},P_{S_i},Q_{S_i}$.

In this study, we adopt a cost function defined as the sum of the construction cost $J_{\rm c}$ and the cumulative long-term maintenance cost $J_{\rm m}$ of edges, i.e., $J=J_{\rm c}+J_{\rm m}$, as in~\cite{Cruz2021Model}.
The construction cost is defined as $J_{\rm c}:=\sum_{(ij)\in\Omega_{\rm L}}c_{\rm c}n_{ij}l_{ij}$ where $c_{\rm c}$ and $l_{ij}$ denote the construction cost of edge per unit length and the length of edge $(ij)$, respectively.
The maintenance cost is defined as $J_{\rm m}:=\delta(\gamma,T)\sum_{(ij)\in\Omega_{\rm L}}c_{\rm m}n_{ij}l_{ij}$ where $c_{\rm m}$ denotes the maintenance cost of the conductor corresponding to the edge per unit length.
The term $\delta(\gamma,T):=\sum_{t_{\rm year}=1}^T 1/(1+\gamma)^{t_{\rm year}-1}$ denotes the present value factor with the annual interest rate $\gamma$ and the number of opeartional years $T$.
Note that it is possible to include the other costs such as power loss costs considered in the existing studies~\cite{Lavorato2012Imposing}.

The distinct feature of our formulation is the existence of the security constraint~\eqref{eq:sec_con}, which makes the problem challenging.
First, it is a bi-level problem that involves supremum with respect to the attack signal $a_i$.
Second, the formulation is infinite-dimensional as the inner maximization with respect to the attack signal $a_i$ is performed over a function space $\mathcal{A}$.
This reflects the generality of our threat model, which accommodates arbitrary temporal profiles for attack signals.
This combination of bi-level structure and infinite-dimensionality highlights intractability by direct solution methods and motivates the need for reformulation.
In the subsequent sections, we propose an equivalent transformation into a tractable form in single-level MILP.

\section{Characterization of Maximally Disruptive Power Injection Signal}
\label{sec:assessment}

\subsection{Maximally disruptive power injection signal characterization problem}
As mentioned in the previous section, the problem is challenging because the search space is an infinite-dimensional function space.
The objective of this section is to characterize the maximally disruptive power injection signal to reduce the problem into a finite-dimensional problem.
Throughout this section, we fix the distribution network structure and the targeted node $i$.
Then we pose the maximally disruptive power injection signal characterization problem as a subproblem as follows:
\begin{equation}
\begin{array}{ll}
    \sup_{a_i\in\mathcal{A}} & \|y_i(\cdot\,;a_i)\|_{\infty}\\
    {\rm s.t.} &  P_j^{\rm a} - \sum_{k\in\Omega_{\rm N}}P_{jk}=0\ \forall j\in\Omega_{\rm L} \\
    &  Q_j^{\rm a}-\sum_{k\in\Omega_{\rm N}}Q_{jk}=0\ \forall j\in\Omega_{\rm L}\\
    &  P_{S_j}-\sum_{k\in\Omega_{\rm N}}n_{jk}P_{jk}=0\ \forall j\in\Omega_{\rm S}\\
    &  Q_{S_j}-\sum_{k\in\Omega_{\rm N}}n_{jk}Q_{jk}=0\ \forall j\in\Omega_{\rm S}\\
    &  P_{jk}=0\ \forall(jk)\in\{(jk)\in\Omega_{\rm E}:n_{jk}=0\}\\
    &  Q_{jk}=0\ \forall(jk)\in\{(jk)\in\Omega_{\rm E} :n_{jk}=0\}\\
    &  v^2_j-v^2_k=2(R_{jk}P_{jk}+X_{jk}Q_{jk})\ \forall (jk)\in\Omega_{\rm E}\\
    &  \dot{Q}_{G_j}(t)=-K_i(v^2_j(t)-v^2_0(t))\ \forall j\in\Omega_{\rm L}.
\end{array}
\end{equation}
Without loss of generality, we can assume that the underlying graph is a spanning tree and only a single substation exists because it suffices to focus only on the subgrid which the targeted node $i$ belongs to.
We provide an analytical solution to the subproblem.

\emph{Remark:} The existing work~\cite{Lindstrom2021PowerInjection} considers the problem with a finite horizon for line grids where all nodes are aligned linearly.
Our analysis in this section extends the results to the infinite time horizon problem for grids with radial topology.

\subsection{State-space representation of network dynamics}
\label{subsec:attackmodel}

We now define the reactance matrix over the entire grid $X$, whose $(i,j)$ entry represents twice the sum of line reactances over the common edges of the paths from the substation to nodes $i$ and $j$, i.e.,
\[
 \textstyle{X(i,j) := 2 \sum_{(kl)\in\mathcal{P}_i\cap\mathcal{P}_j}X_{kl},}
\]
where $\mathcal{P}_i$ denotes the set of edges of the path from the connected substation to node $i$.
We also define the resistance matrix $R$ in a similar manner.
Then the entire dynamics can be expressed in the compact form
\begin{equation}\label{eq:sys_all}
\left\{
\begin{array}{cl}
    \dot{{\bm Q}}_{G}&=-K{\bm y}\\
    {\bm y}&=R({\bm P}_S+{\bm P}_G-{\bm P}_D+{\bm a}^p)\\
    & \quad +X({\bm Q}_S+{\bm Q}_G-{\bm Q}_D+{\bm a}^q)
\end{array}
\right.
\end{equation}
where each bold symbol represents the column vector where the corresponding elements are concatenated vertically and $K$ denotes the diagonal matrix whose $(i,i)$ entry is $K_i$.
We consider the dynamics around the equilibrium $({\bm Q}_{G_e},{\bm y}_e)$ under the condition ${\bm a}=0$.
Take the deviations from the equilibrium $\hat{{\bm Q}}_G:={\bm Q}_G-{{\bm Q}_G}_e$ and $\hat{{\bm y}}:={\bm y}-{\bm y}_e$ as the new variables.
Note that ${\bm y}_e=0$ because of the servomechanism in the inverter, and hence $\hat{\bm y}={\bm y}$.
Furthermore, to simplify the analysis, we make the following assumption consistent with the prior literature~\cite{Bolognani2013Identification}:
\begin{assum}\label{assum:RXratio}
All power lines have the same resistance-to-reactance ratio, i.e., $R_{jk}/X_{jk}=m$ for any $(ij)\in\Omega_{\rm E}$ with some constant $m>0$.
\end{assum}
Assumption~\ref{assum:RXratio} implies that the network uses homogeneous conductor types across feeders.
These lines have similar physical properties, leading to relatively consistent ratios.
Under Assumption~\ref{assum:RXratio}, we have $R=mX$.
By using this relationship, we can merge the two input terms in~\eqref{eq:sys_all} into a single input term as $R{\bm a}^p+X{\bm a}^q=X\hat{\bm a}$ with the surrogate attack signal $\hat{\bm a}:=m{\bm a}^{\rm p}+{\bm a}^{\rm q}$.
This surrogate attack signal has a desirable property that the constraint on $|a_i(t)|$ can be equivalently transformed into that on $|\hat{a}_i(t)|$.
\begin{lemma}\label{lem:attack_signal_equivalence}
    If the original attack signal satisfies $|a_i(t)|\leq C$ the surrogate attack signal satisfies $|\hat{a}_i(t)|\leq C_0$
    where $C_0:=C\sqrt{m^2+1}$.
    Furthermore, if $|\hat{a}_i(t)|\leq C_0$, then there exists an original signal such that $|a_i(t)|\leq C$.
    \end{lemma}
    \begin{proof}
    We omit the variable $t$ for simplicity in this proof.
    Assume that $|a_i|=|a^p_i+{\rm j}a^q_i|\leq C$, which implies that $(a^p_i)^2+(a^q_i)^2\leq C^2$.
    Then we have $\hat{a}^2_i= ((a^p_i)^2+(a^q_i)^2)(m^2+1)-(a^p_i-ma^q_i)^2\leq C^2(m^2+1) - (a^p_i-ma^q_i)^2 \leq C^2(m^2+1)$,
    which leads to the first claim.
    For the second claim, assume that $|\hat{a}_i|\leq C_0$.
    Take $a^p_i=m|\hat{a}_i|/(m^2+1)$ and $a^q_i=|\hat{a}_i|/(m^2+1)$.
    Then $|ma^p_i+a^q_i|=|\hat{a}_i|$ and $(a^p_i)^2+(a^q_i)^2=|\hat{a}_i|^2/(m^2+1)\leq C_0^2/(m^2+1)=C^2$.
\end{proof}
The state-space representation of the grid dynamics with $\hat{\bm a}$ is given by
\begin{equation}\label{eq:ss}
    \left\{
    \begin{array}{cl}
        \dot{\hat{{\bm Q}}}_G&=-KX\hat{{\bm Q}}_G-KX\hat{{\bm a}} \\
        {\bm y}&=X\hat{{\bm Q}}_G+X\hat{{\bm a}}.
    \end{array}
    \right.
\end{equation}
This state-space representation and Lemma~\ref{lem:attack_signal_equivalence} imply that it suffices to consider the problem
\begin{equation}
\textstyle{
\sup_{\hat{a}_i}\ \|y_i(\cdot\,;\hat{a}_i)\|_{\infty}\ {\rm s.t.}\ \eqref{eq:ss},\ |\hat{a}_i(t)|\leq C_0
}
\end{equation}
to characterize the maximally disruptive power injection signal.
Note that all elements but the $i$th component of $\hat{\bm a}$ in~\eqref{eq:ss} take zero constantly.

\subsection{Characterization of maximally disruptive power injection signal}
We further transform the obtained state-space representation to exploit symmetry by the variable transformation ${\bm z}:=K^{-\frac{1}{2}}\hat{{\bm Q}}_G$.
The state-space representation with respect to ${\bm z}$ is given by
\begin{equation}\label{eq:sys_all2}
\left\{
\begin{aligned}
    \dot{{\bm z}}&=-K^{\frac{1}{2}}XK^{\frac{1}{2}}
 {\bm z}-K^{\frac{1}{2}}X\hat{{\bm a}}\\
    {\bm y}&=XK^{\frac{1}{2}}{\bm z}+X\hat{{\bm a}}.
\end{aligned}
\right.
\end{equation}
Letting $A:=K^{\frac{1}{2}}XK^{\frac{1}{2}}$ and $B:=K^{\frac{1}{2}}X$, we can express $|y_i(t\,;\hat{a}_i)|$ as follows:
\begin{equation}
    \label{eq:haty}
    |y_i(t\,;\hat{a}_i)|=\left|{\bm e}^\top_iX{\bm e}_i\hat{a}_i(t)-\int_0^t(B{\bm e}_i)^\top e^{-A(t-\tau)}(B{\bm e}_i)\hat{a}_i(\tau)d\tau\right|.
\end{equation}

The following property on the matrix $X$ is known.
\begin{lemma}
\label{lem:X}
The matrix $X$ is positive definite~\cite{Farivar2013Equilibrium}.
\end{lemma}

Further, we have the following lemma.
\begin{lemma}
 \label{lem:A}
The matrix $e^{-A(t-\tau)}$ is positive definite.
\end{lemma}
\begin{proof}
Since $K$ is a diagonal matrix, $A$ is a symmetric matrix. This implies that $-A(t-\tau)$ is a symmetric matrix.
Because the exponential matrix of a symmetric matrix is positive definite, the claim holds.
\end{proof}

Lemmas \ref{lem:X} and \ref{lem:A} imply that the coefficients in the equation~\eqref{eq:haty} are positive, i.e.,
\begin{equation}\label{eq:CoePositive}
    {\bm e}_i^\top X{\bm e}_i>0,\quad (B{\bm e}_i)^\top e^{-A(t-\tau)}(B{\bm e}_i)>0
\end{equation}
where ${\bm e}_i$ denotes the $i$-th canonical basis.
From those properties, we can characterize the finite-time horizon maximally disruptive power injection signal.
\begin{lemma}
\label{lem:t1t2}
For any $t>0$,
\begin{equation}\label{eq:y_max}
    |y_i(t)|\leq 2{\bm e}_i^\top X{\bm e}_iC_0-{\bm e}_i^\top K^{-\frac{1}{2}}e^{-At}B{\bm e}_iC_0 \left(=: y^{\rm max}_i(t)\right)
\end{equation}
and $\hat{a}_i$ that achieves this maximum is given by
\begin{equation}\label{eq:a_max}
\hat{a}_i(\tau)=
\left\{
\begin{array}{ll}
 \mp C_0, & \tau\in[0,t),\\
 \pm C_0, & \tau=t,\\
 {\rm arbitrary}, & \tau>t.
\end{array}
\right.
\end{equation}
Furthermore, if $0<t_1<t_2$ then $y^{\rm max}_i(t_1)<y^{\rm max}_i(t_2)$.
\end{lemma}
\begin{proof}
From~\eqref{eq:CoePositive}, it is clear that the surrogate attack signal that maximize $|y_i(t)|$ is given by~\eqref{eq:a_max}.
Then $y^{\rm max}$ is given by the right-hand side of~\eqref{eq:y_max}.

The gap between $y^{\rm max}(t_1)$ and $y^{\rm max}(t_2)$ is given by
\begin{equation}
 \label{eq:t1t2}
y^{\rm max}(t_1)-y^{\rm max}(t_2)={\bm e}_i^\top K^{-\frac{1}{2}}(e^{-At_2}-e^{-At_1})B{\bm e}_iC_0.
\end{equation}
From Lemma~\ref{lem:X}, there exists a Cholesky decomposition of $X$ such that $X=LL^\top$ with a lower-triangle matrix $L$~\cite[Corollary~7.2.9]{Horn1985Matrix}.
By using this decomposition, we have
\begin{equation}
\begin{array}{l}
 K^{-\frac{1}{2}}(e^{-At_2}-e^{-At_1})B\\
 =X\left\{\sum_{k=0}^\infty\frac{(-KXt_2)^k}{k!}-\sum_{k=0}^\infty\frac{(-KXt_1)^k}{k!}\right\}\\
 =L(e^{-L^\top KLt_2}-e^{-L^\top KLt_1})L^\top.
\end{array}
\end{equation}
Because $L^\top KL$ is positive definite, $e^{-L^\top K L t}$ is also positive definite.
Hence there exists an orthogonal matrix $U$ such that
$Ue^{-L^\top K L t}U^\top = {\rm diag}(e^{-\mu_1t},\ldots,e^{-\mu_Nt})$
where $\mu_1,\ldots,\mu_N$ are positive eigenvalues of $L^\top KL$.
Therefore,
$e^{-L^\top KLt_2}-e^{-L^\top KLt_1}=U^\top{\rm diag}(e^{-\mu_1t_2}-e^{-\mu_1t_1},\ldots,e^{-\lambda'_Nt_2}-e^{-\mu_Nt_1})U.$
Since $t_1<t_2$, we have $e^{-\mu_k t_2} - e^{-\mu_k t_1}<0$ for any $k=1,\ldots,N$.
Thus ${\bm w}^\top(e^{-L^\top KLt_2}-e^{-L^\top KLt_1}){\bm w}<0$, which means  ${\bm w}^\top K^{-\frac{1}{2}}(e^{-At_2}-e^{-At_1})B{\bm w}<0$, for any ${\bm w}\neq 0$.
Then \eqref{eq:t1t2} implies $y^{\rm max}(t_1)-y^{\rm max}(t_2)<0$.
\end{proof}

\begin{figure}
    \centering
    \includegraphics[width=0.98\linewidth]{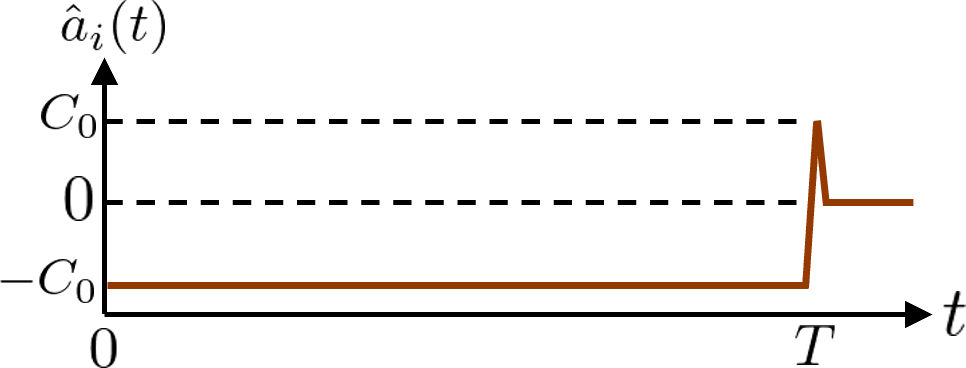}
    \caption{The Maximally disruptive power injection signal.}
    \label{fig:attack-signal}
\end{figure}

The rationale of the Maximally disruptive power injection signal for the finite time horizon problem is given as follows.
Due to the monotonicity of the distribution system, the reactive power evolves monotonically to compensate the injected power that constantly takes the maximum power for $[0,t)$.
Flipping the power in the opposite direction at the terminal time instant $t$ detrimentally affects the transitioned reactive power.
Furthermore, the attacker would inject the malicious power as long as possible since its impact is monotonically increasing.
The temporal profile of the Maximally disruptive power injection signal is illustrated in Fig.~\ref{fig:attack-signal}.

The following theorem that characterizes the maximally disruptive power injection signal is the main result of this section.
\begin{theorem}
 \label{the:worstattack-complete}
The supremum of the maximally disruptive power injection signal characterization problem is given by
\begin{equation}
\textstyle{2{\bm e}_i^\top X{\bm e}_iC_0=2C_0\sum_{(jk)\in\mathcal{P}_{i}}X_{jk}}.
\end{equation}
\end{theorem}
\begin{proof}
From Lemma~\ref{lem:t1t2}, the supremum is given by $\lim_{t\to+\infty}y^{\rm max}(t) = 2{\bm e}_i^\top X{\bm e}_iC_0$.
The form of the matrix $X$ leads to the claim.
\end{proof}

The obtained characterization indicates that the higher the total reactances of the lines on the path from the substation to the targeted node is the worse the attack impact is.
Hence, to evaluate the effect of the maximally disruptive power injection signal, we no longer need to consider the infinite-dimensional function space.
Instead, we only have to consider a finite-dimensional discrete space related to the path from the substation to the targeted node.
It should be emphasized that this result agrees with the existing analysis on voltage volatility for uncertain renewable distributed generation~\cite{Song2019New}.
Note also that this finding implies that the attacker who aims at maximizing the attack impact selects the node with the largest reactance on the path from the substation as the target node.

\section{Equivalent Reformulation to MILP}
\label{sec:milp}

Theorem~\ref{the:worstattack-complete} indicates that, when the attack signal is chosen in this maximization form, the resulting impact simplifies to $\|y_i(\cdot\,;a_i)\|_{\infty} = 2C_0\sum_{(jk)\in\mathcal{P}_i}X_{jk}$, where the expression is fully determined by network parameters and path structure.
This characterization removes the infinite-dimensionality from the inner maximization problem, allowing our security constraint~\eqref{eq:sec_con} to be replaced with an equivalent constraint
\begin{equation}\label{con:path}
\textstyle{
2C_0\sum_{(jk)\in\mathcal{P}_{i}}X_{jk}\leq \overline{y}\quad  \forall i\in\Omega_{\rm L}.
}
\end{equation}
Consequently, the original bi-level problem is reduced to a tractable finite-dimensional optimization problem.
However, the constraint in this form is still difficult to handle because of the existence of the path $\mathcal{P}_i$, which cannot be directly expressed in terms of the existing decision variables.
Despite being finite-dimensional, the path-dependent structure still leads to a general nonlinear integer program that cannot be solved directly using standard solvers.
The key observation to resolve the issue is that this constraint can be regarded as the \emph{bottleneck shortest path problem}~\cite{Turner2011Variants}, which deals with finding structures that minimize the maximum of the path weights.
We employ a well-known technique for shortest path problems to transform it into an MILP~\cite{Rafael2016Formulations}.

Specifically, we introduce auxiliary binary decision variables $x^i_{jk}$, which are set to $1$ if edge $(jk)$ lies on the path $\mathcal{P}_i$ and $0$ otherwise.
Under this definition, we observe that $\sum_{(jk)\in\mathcal{P}_i}X_{jk} = \sum_{(jk)\in\Omega_{\rm E}}x^i_{jk}X_{jk}$, which is linear with respect to $x^i_{jk}$.
The variables are subject to the following constraints.
\begin{subequations}\label{eq:e_P}
\begin{align}
& x^i_{jk}\in\{0,1\},\quad \forall i\in\Omega_{\rm L},\forall(jk)\in\Omega_{\rm E},\label{eq:e_Pbinary}\\
& x^i_{jk}\leq n_{jk},\quad \forall i\in\Omega_{\rm L},\forall(jk)\in\Omega_{\rm E},\label{eq:e_Pedge}\\
& \sum_{j:(sj)\in\Omega_{\rm E}} x^i_{sj} - \sum_{j:(js)\in\Omega_{\rm E}} x^i_{js}=1,\quad \forall i\in\Omega_{\rm L},\label{eq:e_Psub}\\
& \sum_{j:(ij)\in\Omega_{\rm E}} x^i_{ij} - \sum_{j:(ji)\in\Omega_{\rm E}} x^i_{ji}=-1,\quad \forall i\in\Omega_{\rm L}\label{eq:e_Pi},\\
& \sum_{k:(kj)\in\Omega_{\rm E}} x^i_{kj} - \sum_{l:(jk)\in\Omega_{\rm E}}x^i_{jl}=0,\quad \forall j\in\Omega_{\rm L}\setminus\{i\}\label{eq:e_Pkl}.
\end{align}
\end{subequations}
These constraints collectively ensure that the variables $x^i_{jk}$ accurately represent the edges on the path $\mathcal{P}_i$.
Specifically, the first constraint~\eqref{eq:e_Pbinary} enforces the binary nature of the variables.
The second constraint~\eqref{eq:e_Pedge} ensures that an edge may only be part of the path if it is constructed, i.e., $x^i_{jk}=0$ if $n_{jk}=0$.
The third constraint~\eqref{eq:e_Psub} guarantees that exactly one edge departs from the substation along the path.
The fourth constraint~\eqref{eq:e_Pi} ensures that exactly one edge enters node $i$.
The fifth constraint~\eqref{eq:e_Pkl} enforces flow conservation at all intermediate nodes on the path.

\emph{Example:}
Consider the planning shown in Fig.~\ref{fig:ex-design}, where the path $\mathcal{P}_3$ consists of edges $(S_14)$ and $(43)$.
In this case, $x^3_{S_14}=x^3_{43}=1$ and all other $x^3_{jk}=0$.
It is straightforward to verify that all constraints are satisfied.
First, the binary and edge-existence constraints~\eqref{eq:e_Pbinary} and \eqref{eq:e_Pedge} hold trivially.
Second, the substation constraint~\eqref{eq:e_Psub} is satisfied since only one edge originates from $S_1$ ($x^3_{S_14}=1$).
Third, the destination constraint~\eqref{eq:e_Pi} holds due to the inclusion of edge $(4,3)$ ($x^3_{43}=1$).
Finally, the flow conservation constraint~\eqref{eq:e_Pkl} is met since $x^3_{S_14}-x^3_{43}=0$

By using the auxiliary variables, the original inequality~\eqref{con:path} can be equivalently rewritten as
\[
 2C_0 \sum_{(jk)\in\Omega_{\rm E}}x^i_{jk}X_{jk}\leq \bar{y},\quad\forall i\in\Omega_{\rm L},
\]
where the summation over the path $\mathcal{P}_i$ is expressed using the auxiliary binary variables $x^i_{jk}$.
Combined with the constraints~\eqref{eq:e_P}, which encode the structure of the path $\mathcal{P}_i$ using linear equalities and inequalities, the entire formulation remains to be an MILP.
After all, the resilient distribution network planning problem can now equivalently be transformed into
\begin{align}
    \min \quad & J \notag \\
    {\rm s.t.}\quad & \eqref{eq:oneway}\textrm{-}\eqref{eq:Qconstraint}, \eqref{eq:radiality},\eqref{eq:e_P} \\
    &\textstyle 2C_0\sum_{(jk)\in\Omega_{\rm E}}x^i_{jk}X_{jk} \leq \overline{y} && \forall i\in\Omega_{\rm L},
\end{align}
with decision variables $n_{ij},P_{ij},Q_{ij},P_{S_i},Q_{S_i},x^i_{jk}$.
Note that, because the temporal pattern of the attack signal is fixed in the reformulated constraint, the inner problem no longer involves a maximization over a function space, but instead reduces to a deterministic inequality constraint.
The transformed problem is now an MILP, which is tractable by using existing solvers.

\section{Numerical Simulation}
\label{sec:simulation}

\subsection{Simulation settings}
We verify the effectiveness of the proposed method through numerical simulations.
We use the MILP solver provided by the IBM CPLEX Optimization Studio 22.1.1~\cite{IBM2022}.
The CPU used for the simulations is Intel Xeon Bronze 3104 CPU @ 1.70GHz (2 processors) with 32.0 GB RAM.

\begin{figure*}
\centerline{
\subfloat[Underlying network of the grid used in the simulation.]{
\includegraphics[clip, width=0.25\textwidth]{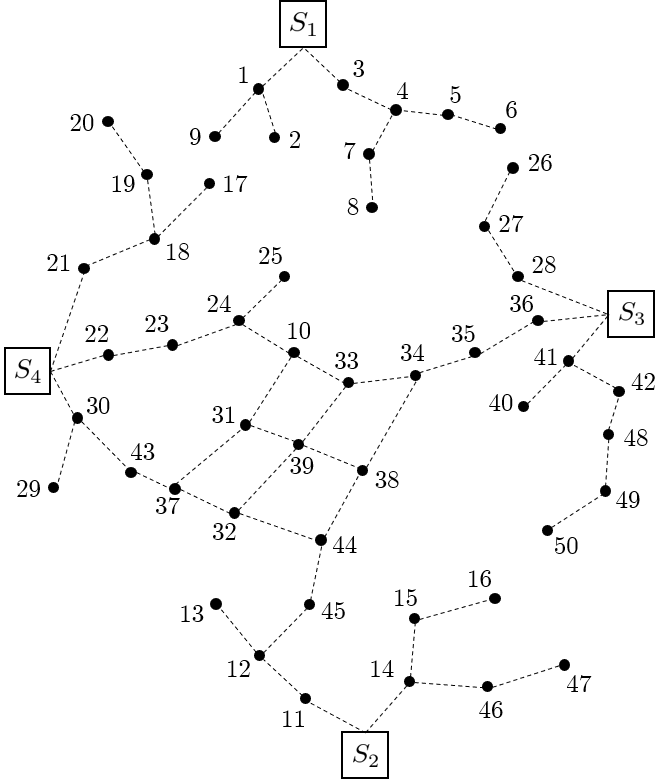}
\label{fig:network-for-simulation}
}
\hfil
\subfloat[Planned network without security constraints.]{
\includegraphics[clip, width=0.25\textwidth]{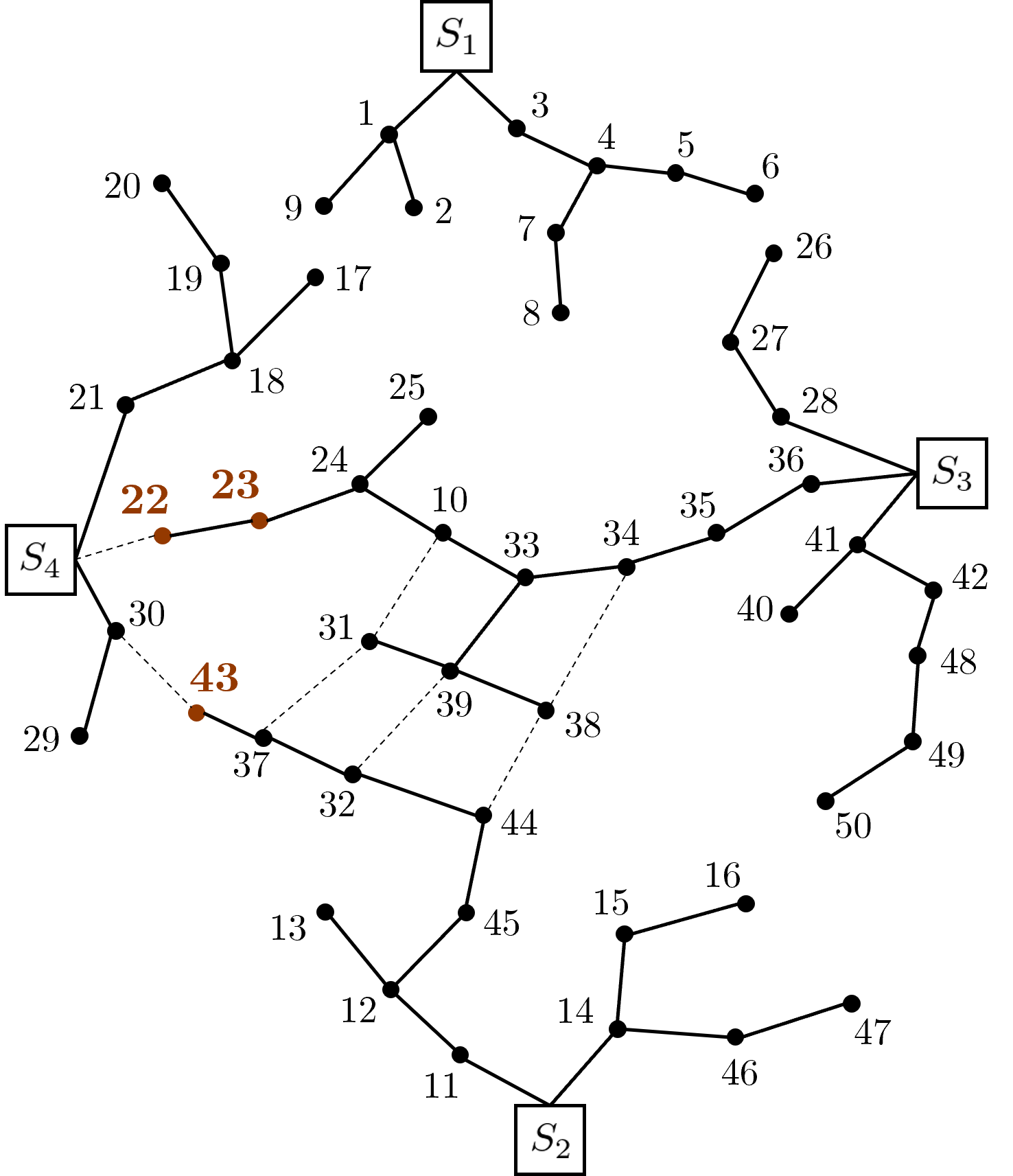}
\label{fig:cost}
}
\hfil
\subfloat[Planned network with security constraints.]{
\includegraphics[clip, width=0.25\textwidth]{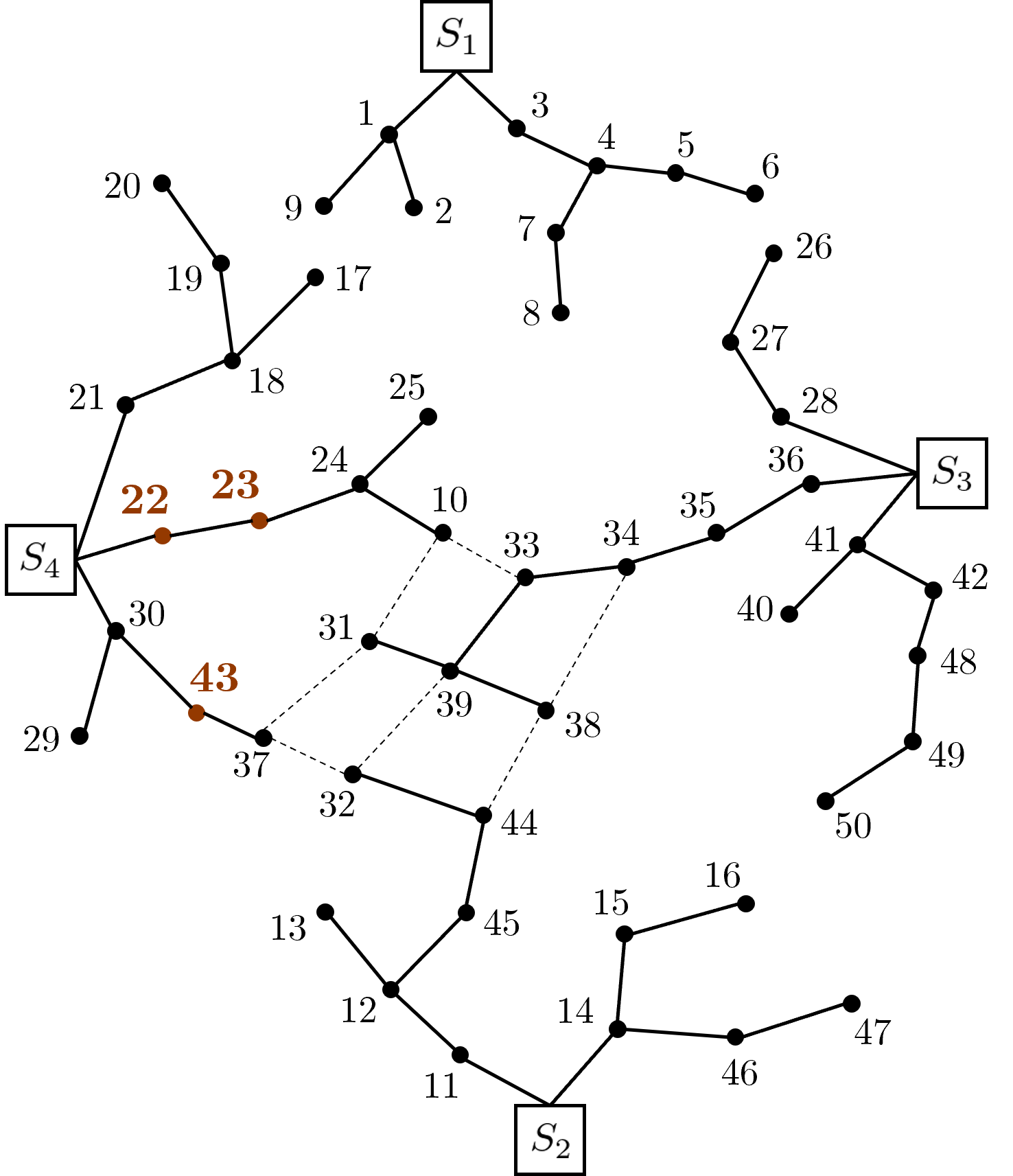}
\label{fig:security}
}
}
\caption{Simulated distribution network.}
\label{fig:simulated-network}
\end{figure*}

The network data is a modified version of the one used in \cite{Cruz2021Model,Lavorato2012Imposing,Miranda1994GA}.
The underlying topology of the network is shown in Fig.~\ref{fig:network-for-simulation}, which includes 4 substations, 50 consumer nodes, and 57 expandable lines.
The demand at each consumer node~\cite{Goncalves2015Shortterm} is shown in Table~\ref{tab:node}.
Each consumer is assumed to cover 30\% of its demand through self-generation using PV.
The rated voltage is 15~kV, and the tolerable voltage range is $\pm 5\%$.
The resistance and reactance of the edges depend on their length and are set to 0.3655 $\Omega/\text{km}$ and 0.2520 $\Omega/\text{km}$, respectively.
The large constant is set as $M=60,000$~kW.
The construction cost of each edge is $5,000$~US\$/km.
The annual maintenance cost of each edge is $450$~US\$/km with the annual interest rate $\gamma=0.1$ and the operational years $T=10$.
The length of each edge is given in Table~\ref{tab:edge-length} in units of kilometers~\cite{Cruz2021Model}.

\begin{table}[]
    \centering
    \caption{Consumer Data}
    \begin{tabular}{ccc@{\hspace{20pt}}ccc}\hline
        $i$ & $P_{D_i}$~[kW] & $Q_{D_i}$~[kVAr] & $i$ & $P_{D_i}$ & $Q_{D_i}$ \\ \hline
        1 & 3622.50 & 2012.50 & 26 & 1035.00 & 575.00 \\
        2 & 1293.75 & 718.75 & 27 & 1293.75 & 718.75 \\
        3 & 603.75 & 335.42 & 28 & 603.75 & 335.42 \\
        4 & 948.75 & 527.08 & 29 & 1207.50 & 670.83 \\
        5 & 2242.50 & 1245.83 & 30 & 2242.50 & 1245.83 \\
        6 & 603.75 & 335.42 & 31 & 603.75 & 335.42 \\
        7 & 862.50 & 479.17 & 32 & 1466.25 & 814.58 \\
        8 & 1638.75 & 910.42 & 33 & 2501.25 & 1389.58 \\
        9 & 1035.00 & 575.00 & 34 & 1035.00 & 575.00 \\
        10 & 2501.25 & 1389.58 & 35 & 776.25 & 431.25 \\
        11 & 258.75 & 143.75 & 36 & 258.75 & 143.75 \\
        12 & 1552.50 & 862.50 & 37 & 1811.25 & 1006.25 \\
        13 & 948.75 & 527.08 & 38 & 948.75 & 527.08 \\
        14 & 862.50 & 479.17 & 39 & 862.50 & 479.17 \\
        15 & 1207.50 & 670.83 & 40 & 127.50 & 670.83 \\
        16 & 1638.75 & 910.42 & 41 & 776.25 & 431.25 \\
        17 & 603.75 & 335.42 & 42 & 1035.00 & 575.00 \\
        18 & 1035.00 & 575.00 & 43 & 1121.25 & 622.92 \\
        19 & 1207.50 & 670.83 & 44 & 1207.50 & 670.83 \\
        20 & 690.00 & 383.33 & 45 & 690.00 & 383.33 \\
        21 & 1552.50 & 862.50 & 46 & 1552.50 & 862.50 \\
        22 & 948.75 & 527.08 & 47 & 862.50 & 479.17 \\
        23 & 862.50 & 479.17 & 48 & 690.00 & 383.33 \\
        24 & 431.25 & 239.58 & 49 & 431.25 & 239.58 \\
        25 & 776.25 & 431.25 & 50 & 690.00 & 383.33 \\ \hline
    \end{tabular}
    \label{tab:node}
\end{table}

\begin{table}[]
    \centering
    \caption{Edge Data}
    \begin{tabular}{ccc@{\hspace{20pt}}ccc@{\hspace{20pt}}ccc} \hline
        $s$ & $g$ & $l_{sg}$ [km] & $s$ & $g$ & $l_{sg}$ & $s$ & $g$ & $l_{sg}$ \\ \hline
        $S_1$ & 1 & 1.44 & 38 & 39 & 1.23 & 48 & 49 & 1.92 \\
        1 & 2 & 1.60 & 14 & 46 & 1.76 & 49 & 50 & 1.12 \\
        1 & 9 & 1.76 & 46 & 47 & 1.60 & $S_4$ & 21 & 1.28 \\
        $S_1$ & 3 & 1.12 & $S_3$ & 28 & 1.60 & 21 & 18 & 1.60 \\
        3 & 4 & 1.60 & 28 & 27 & 1.60 & 18 & 19 & 1.28 \\
        4 & 5 & 1.60 & 27 & 26 & 1.76 & 19 & 20 & 1.60 \\
        5 & 6 & 1.28 & $S_3$ & 36 & 1.28 & 18 & 17 & 2.08 \\
        4 & 7 & 1.28 & 36 & 35 & 1.12 & $S_4$ & 22 & 1.92 \\
        7 & 8 & 1.60 & 35 & 34 & 1.12 & 22 & 23 & 1.76 \\
        $S_2$ & 11 & 1.44 & 34 & 33 & 0.96 & 23 & 24 & 1.44 \\
        11 & 12 & 1.60 & 33 & 39 & 1.44 & 24 & 25 & 1.12 \\
        12 & 13 & 2.24 & 39 & 31 & 1.23 & 24 & 10 & 1.23 \\
        $S_2$ & 14 & 1.92 & 39 & 32 & 2.08 & 10 & 33 & 1.23 \\
        14 & 15 & 1.92 & 32 & 37 & 1.23 & $S_4$ & 30 & 1.44 \\
        15 & 16 & 1.44 & 32 & 44 & 1.23 & 30 & 29 & 1.60 \\
        12 & 45 & 1,28 & $S_3$ & 41 & 1.60 & 30 & 43 & 2.08 \\
        45 & 44 & 1.12 & 41 & 40 & 1.28 & 43 & 37 & 1.28 \\
        44 & 38 & 1.60 & 41 & 42 & 1.92 & 37 & 31 & 1.92 \\
        38 & 34 & 2.50 & 42 & 48 & 1.28 & 31 & 10 & 1.60 \\ \hline
    \end{tabular}
    \label{tab:edge-length}
\end{table}

\subsection{Simulation results}
First, we plan the distribution network without the security constraint, where there are 350 decision variables, out of which 114 are binary variables, and there are 638 constraints.
The planned network is illustrated in Fig.~\ref{fig:cost}.
The solid and dashed lines represent the edges that are deployed and not, respectively.
It can be confirmed that the network is radial and all consumer nodes are connected to one of the substations.

\begin{figure*}
    \centering
    \includegraphics[width=.9\textwidth]{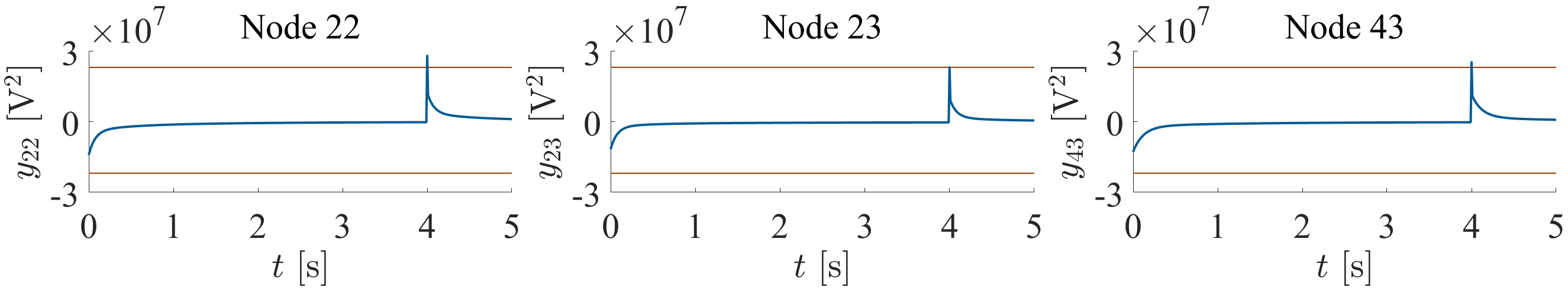}
    \caption{Responses that exceeds the tolerable range against the attack to each node in planned network w/o security constraints.}
    \label{fig:worst3-damage-cost}
\end{figure*}

We consider injecting the maximally disruptive power injection signal characterized in Lemma~\ref{lem:t1t2} with $C_0=5,550,000$ and $t=4$.
Note that the value of $C_0$ corresponds to the attack power 2,594~kW, which is comparable to typical consumer loads (see TABLE~I).
The three most vulnerable nodes in the plan without the security constraint are Nodes 22, 43, and 23, as shown in TABLE~\ref{table:path_reactances}.
Fig.~\ref{fig:cost} illustrates that these nodes are located away from the connected substations.
We consider three distinct scenarios, each corresponding to an attack targeting a single node, either node 22, 23, or 43.
Each response at the corresponding node is shown in Fig.~\ref{fig:worst3-damage-cost}, where the predefined tolerable range is also indicated.
As observed, the voltage deviations exceed the tolerable range in each scenario, indicating that the current planning fails to meet the attack-resiliency requirement.

\begin{figure*}
    \centering
    \includegraphics[width=.9\textwidth]{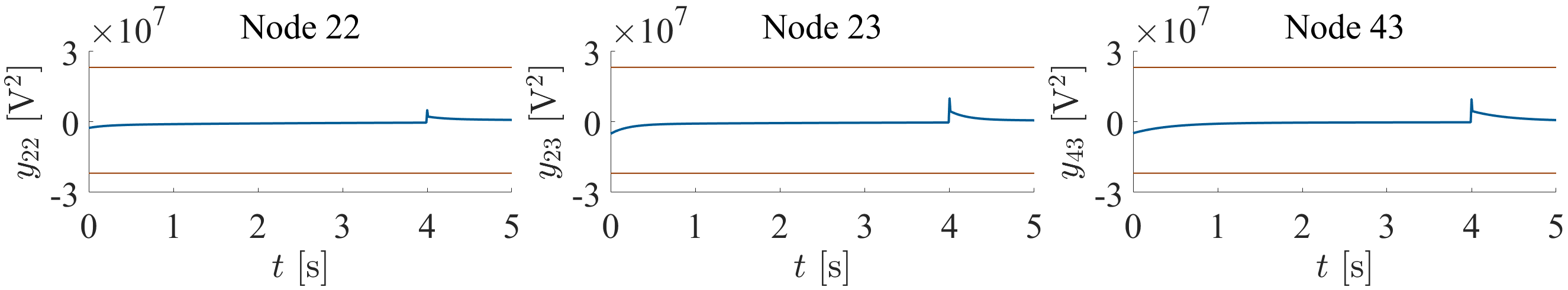}
    \caption{Responses at nodes above against the attack to each node in planned network with security constraints.}
    \label{fig:worst3-damage-security}
\end{figure*}

\begin{table}[t]
\caption{Total path reactances at three most vulnerable nodes.}
\label{table:path_reactances}
\centering
\begin{tabular}{lllllll}
\toprule
 & \multicolumn{3}{c}{w/o security constraint} & \multicolumn{3}{c}{with security constraint}\\
 \cmidrule(lr){2-4} \cmidrule(lr){5-7}
vulnerable node & 22 & 43 & 23 & 50 & 31 & 38\\
path reactance [$\Omega$] & 2.56 & 2.31 & 2.11 & 1.97 & 1.80 & 1.80\\
\bottomrule
\end{tabular}
\end{table}

Next, we plan the distribution network with the security constraint.
In this case, there are 6,050 decision variables, out of which 5,814 are binary variables, and there are 11,838 constraint conditions.
The execution time for solving this problem is 7.899 seconds.
The planned network is illustrated in Fig.~\ref{fig:security}.
Radiality and connectivity to a substation can be confirmed in this case as well.

We consider injecting the maximally disruptive signal to the network.
The responses against attacks to Node 22, 23, and 43 are compared in Fig.~\ref{fig:worst3-damage-security}.
In contrast to the case without the security constraint, the response remains within the tolerable range, because of the close proximity to the connected substation $S_4$ as illustrated in Fig.~\ref{fig:security}.
The three most vulnerable nodes with the security constraint are given in TABLE~\ref{table:path_reactances}, where the most fragile node here is Node 50, but the path reactance is much lower than those in the plan without the security constraint and the maximum voltage deviation at Node 50 is kept within the tolerable range.
This result demonstrates the effectiveness of the proposed planning method.

\begin{figure}
    \centering
    \includegraphics[width=0.9\linewidth]{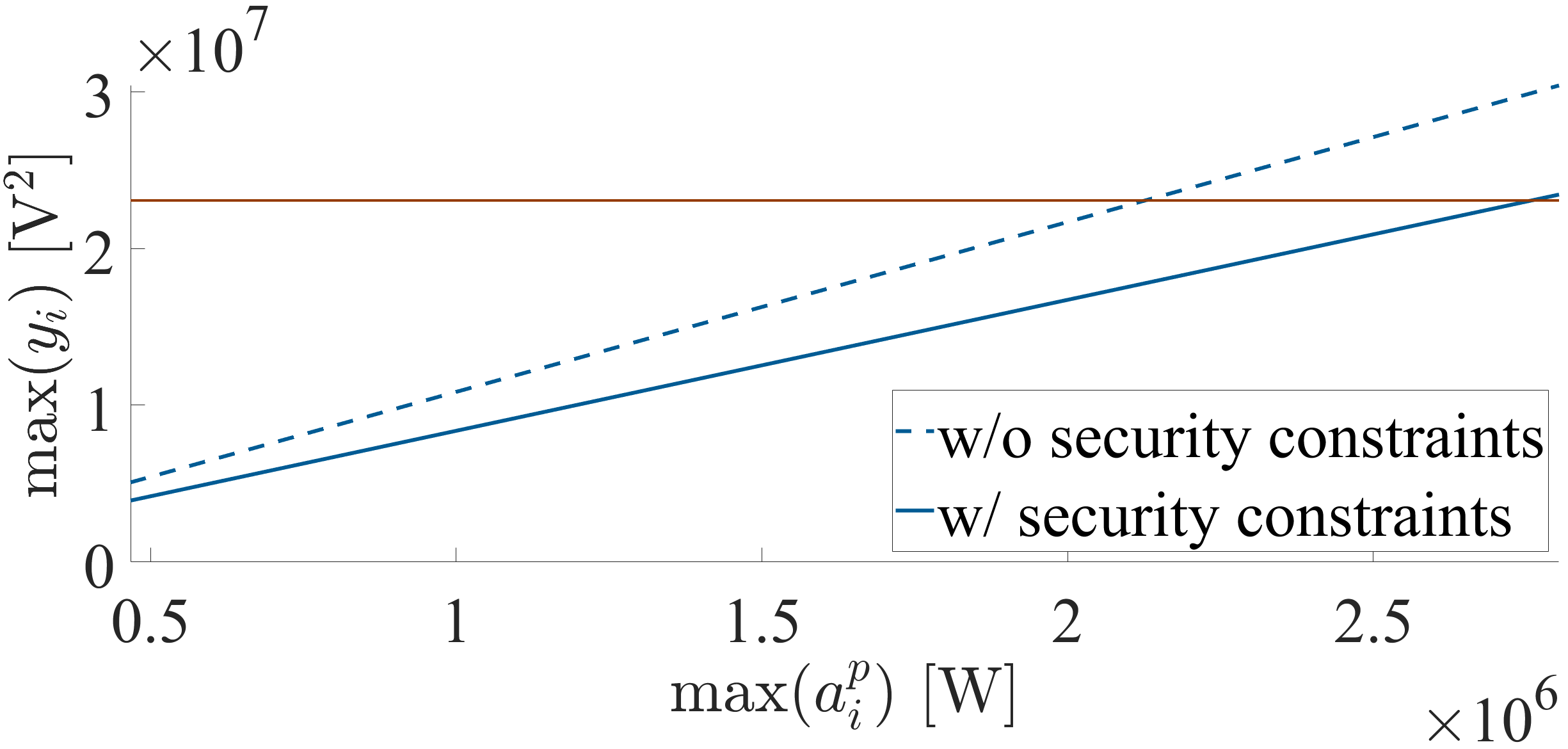}
    \caption{Maximum squared voltage fluctuation as a function of the attacker's budget.}
    \label{fig:compare-max-haty}
\end{figure}

Fig.~\ref{fig:compare-max-haty} illustrates the relationship between the magnitude of the attack power and the maximum value of the response $y_i$ at the most fragile node for the planned network with and without the security constraint, where the red line depicts the tolerable range.
The result indicates that, with security constraints, the solution allows for larger magnitudes of tolerable attack power compared to the solution without security constraints.
Moreover, we observe that as the assumed attack power approaches the threshold at which the voltage fluctuation exceeds the tolerable range, the network topology obtained through the planning with the security constraint transitions from a financially efficient configuration to a more resilient one.
At this point, the most vulnerable node also changes from Node 22 to Node 50.
Note also that the results show a linear relationship, indicating that the proposed method is not highly sensitive to this parameter.

The tolerable attack power, the construction cost, and the maintenance cost for both plans are compared in Table~\ref{tab:compare}.
The tolerable attack power is computed by deriving the maximum tolerable power $C$ from the theorem with $C_0$ determined from the security constraint.
The inclusion of the security constraint result in a marginal increase (approximately 2.1\%) in both construction and maintenance costs.
Instead, the tolerable attack power has increased from 2108.9~kW to 2727.6~kW, approximately a 29.3\% improvement.
It is worth noting that the minimal increase in cost stems from the nature of our resilience enhancement approach, which modifies the placement of conductors without increasing their total number.
Since both the investment and operational costs are primarily determined by the total number of conductors, the overall cost remains largely unchanged.
On the other hand, as revealed by our analysis, resilience is primarily influenced by the placement of conductors rather than their number.
Consequently, substantial improvements in network resilience against adversarial power injection are achieved with only a minimal increase in cost.

\begin{table}
    \centering
    \caption{Comparison of Planned Networks}
    \begin{tabular}{m{0.2\linewidth}m{0.325\linewidth}m{0.325\linewidth}} \hline
         & w/o security constraints & w/ security constraints \\ \hline
        Unused edges&44-38,~38-34,~39-32,~$S_4$-22,~30-43,~37-31,~31-10 & 44-38,~38-34,~39-32,~32-37,~10-33,~37-31,~31-10 \\
        Tolerable attack power & 2108.9~[kW] & 2727.6~[kW] (+29.3\%) \\
        Construction cost & 367300~[US\$] & 375000~[US\$] (+2.1\%) \\
        Maintenance cost & 223470~[US\$] & 228150~[US\$] (+2.1\%) \\\hline
    \end{tabular}
    \label{tab:compare}
\end{table}

\section{Conclusion}
\label{sec:conclusion}
This study has proposed a resilient distribution network planning method against dynamic malicious power injection attacks.
The optimization problem is formulated based on the conventional planning problem with a voltage volatility constraint against dynamic malicious power injection attacks.
To derive a tractable form, we have analytically characterized the maximally disruptive power injection signal.
Additionally, by applying a common technique used in the shortest path problem, we successfully reformulated the planning problem in the form of an MILP.
We have conducted numerical simulations to validate the effectiveness of the proposed design problem.

Future work will focus on four key directions.
The first involves generalizing the problem formulation to incorporate non-uniform characteristics of lines, multiple target nodes, and diverse objective functions.
In particular, integrating existing studies that address general objectives will help validate the significance of the proposed security enhancement.
The second direction is to account for the influence of high-voltage transmission networks, because security failures can have an escalated impact when they coincide with disturbances in the neighboring transmission network.
Thirdly, the assumption on the constant line resistance–reactance ratio limits the generality of the method.
A possible extension is to apply perturbation analysis to account for non-uniform line characteristics, which may allow deriving an upper bound on the voltage deviation error relative to the ideal case.
Finally, the fourth direction involves incorporating fast dynamic factors, such as load variations caused by electric vehicle integration or sudden cloud cover.
We anticipate that stochastic methods will prove effective to manage such fast uncertainty.

\bibliographystyle{IEEEtran}
\bibliography{bibtex/bib/myref}

\begin{IEEEbiography}
[{\includegraphics[width=1in,height=1.25in,clip,keepaspectratio]{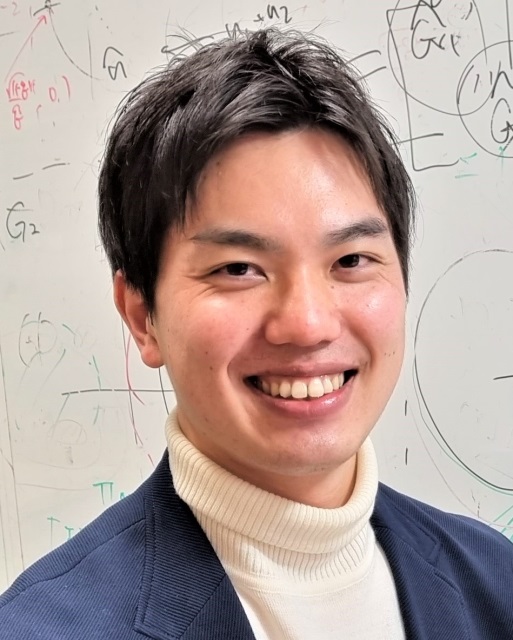}}]{Hampei Sasahara}
(M'19)
is Lecturer with the Department of Information Physics and Computing, Graduate School of Information Science and Technology, the University of Tokyo, Tokyo, Japan.
He received the Ph.D. degree in engineering from Tokyo Institute of Technology in 2019.
From 2019 to 2021, he was a Postdoctoral Scholar with KTH Royal Institute of Technology, Stockholm, Sweden.
From 2022 to 2024, he was an Assistant Professor with Tokyo Institute of Technology, Tokyo, Japan.
From 2024 to 2025, he was an Assistant Professor with Institute of Science Tokyo, Tokyo, Japan.
His main interests include secure control system design and control of large-scale systems.
\end{IEEEbiography}

\begin{IEEEbiography}
[{\includegraphics[width=1in,height=1.25in,clip,keepaspectratio]{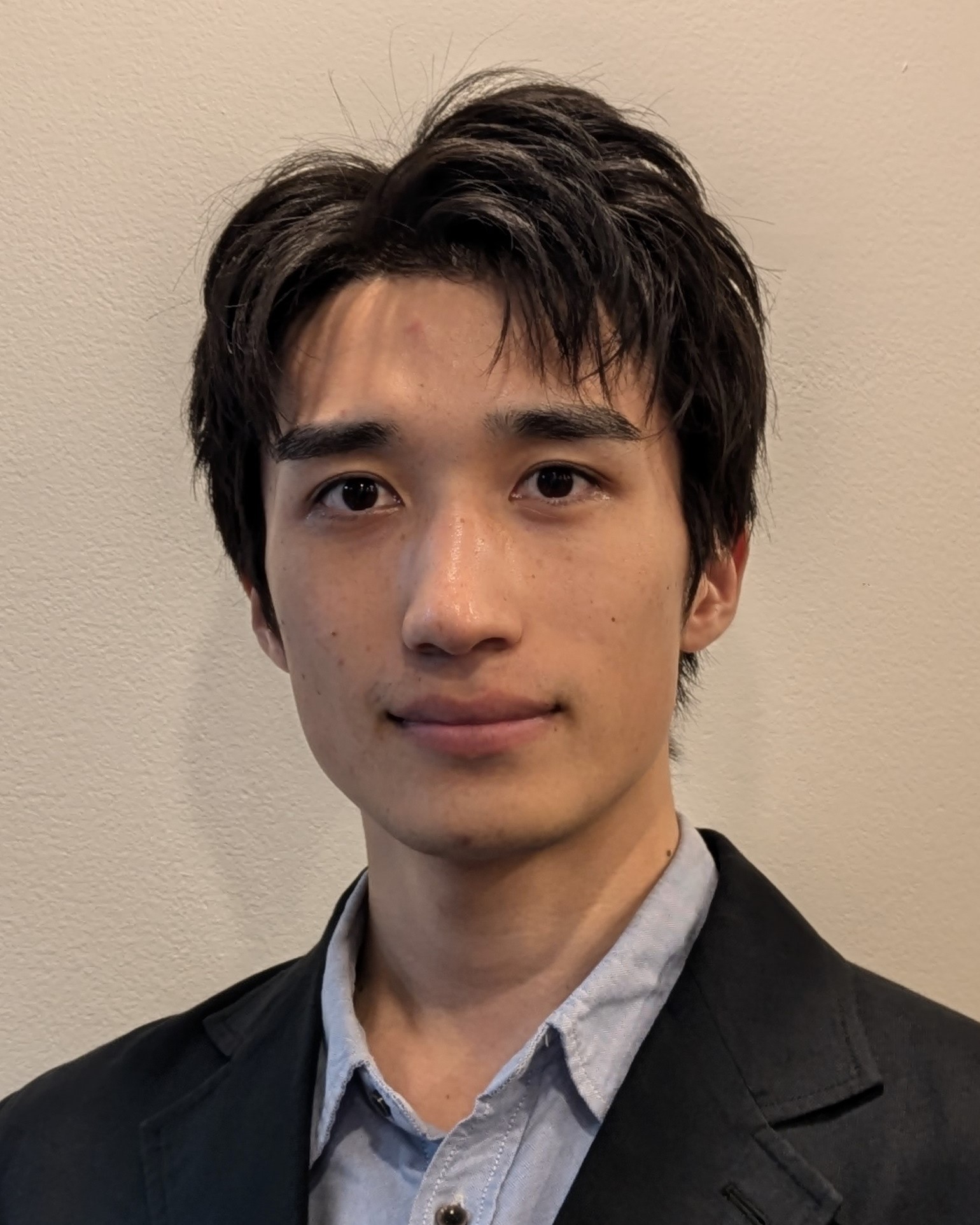}}]{Tatsuya Yamada}
received the Master's degree in engineering from Tokyo Institute of Technology in 2024.
His research interests include security of smart grids.
\end{IEEEbiography}

\begin{IEEEbiography}[{\includegraphics[width=1in,height=1.25in,clip,keepaspectratio]{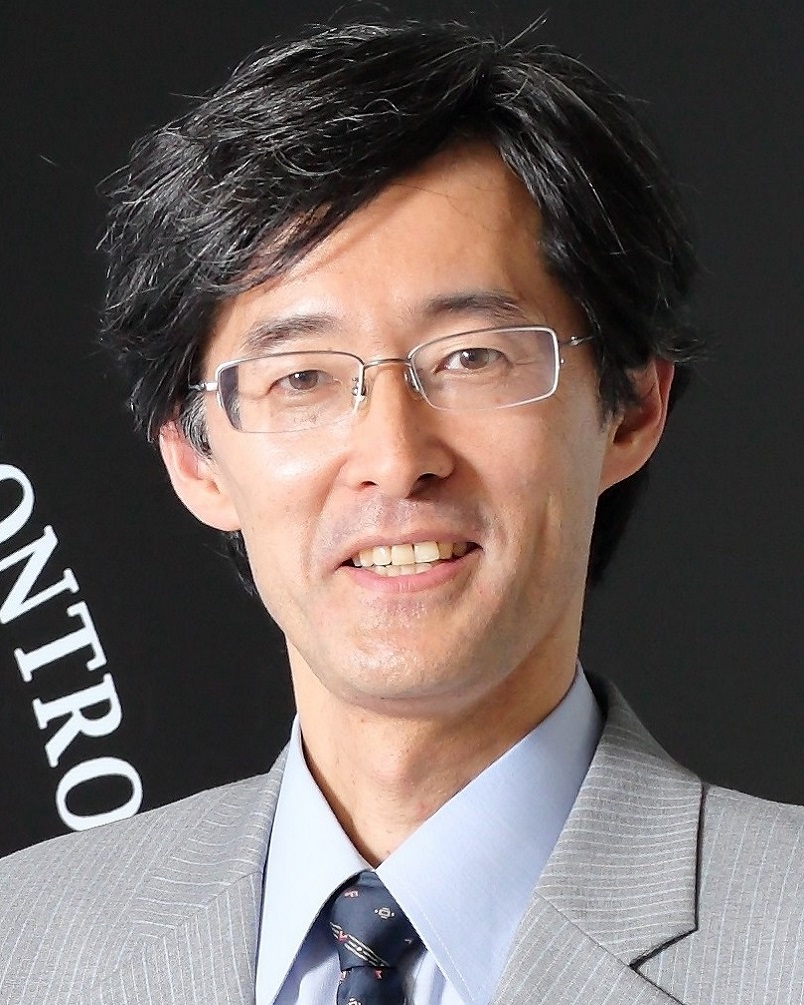}}]{Jun-ichi Imura}(SM’18)
received
the Ph.D. degree in mechanical engineering from Kyoto University, Kyoto, Japan, in 1990 and 1995, respectively.
From 2001 to 2024, he had been with Tokyo Institute of Technology, Tokyo, Japan.
He is currently a Professor at the Department of Systems and Control Engineering, Institute of Science Tokyo.
His research interests include modeling, analysis, and synthesis of nonlinear systems, hybrid systems, and large-scale network systems with applications to power systems, ITS, biological systems, and industrial process systems.
\end{IEEEbiography}

\begin{IEEEbiography}[{\includegraphics[width=1in,height=1.25in,clip,keepaspectratio]
{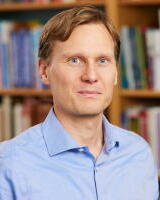}}]{Henrik Sandberg}(M'04-SM'21-F'23)
is Professor at the Division of Decision and Control Systems, KTH Royal Institute of Technology, Stockholm, Sweden.
He received the M.Sc. degree in engineering physics and the Ph.D. degree in automatic control from Lund University, Lund, Sweden, in 1999 and 2004, respectively.
His current research interests include security of cyber-physical systems, power systems, model reduction, and fundamental limitations in control.
\end{IEEEbiography}




\end{document}